\documentclass[longauth]{aa} 
\usepackage[utf8]{inputenc}
\usepackage{graphicx}
\usepackage{siunitx}
\usepackage{amsmath}
\usepackage{textcomp}
\usepackage{multirow}               
\usepackage{booktabs}                             
\usepackage{multicol,lipsum}
\usepackage{colortbl}
\usepackage{makecell} 
\usepackage{float}

\usepackage{txfonts}

\usepackage[]{hyperref}
\usepackage{hyperref}
 \hypersetup{
     colorlinks=true,
     linkcolor=blue,
     filecolor=blue,
     citecolor =blue,      
     urlcolor=blue,
     }

\DeclareRobustCommand{\ion}[2]{%
\relax\ifmmode
\ifx\testbx\f@series
{\mathbf{#1\,\mathsc{#2}}}\else
{\mathrm{#1\,\mathsc{#2}}}\fi
\else\textup{#1\,{\mdseries\textsc{#2}}}%
\fi}

\usepackage{comment}
\usepackage{balance}
\usepackage{subfig}

\usepackage{upgreek}
\newcommand{\micron}{$\upmu$m}
\usepackage{chemformula}
\definecolor{royalblue}{rgb}{0.255, 0.412, 0.882}

\begin{document} 

   \title{The GAPS Programme at TNG\thanks{Based on observations made with the Italian \textit{Telescopio Nazionale Galileo} (TNG), operated on the island of La Palma by the INAF - \textit{Fundaci\'on Galileo Galilei} at the \textit{Roque de Los Muchachos} Observatory of the \textit{Instituto de Astrof\'isica de Canarias} (IAC).}} 
   \subtitle{LXXIII. Confirmation of the hot sub-Neptune TOI-4602\,b (HD 25295 b), a key target for future atmospheric characterization}
   \titlerunning{Confirmation of the hot sub-Neptune TOI-4602\,b}
    \authorrunning{Di Maio et al.}
    
   \author{C. Di Maio
          \inst{\ref{inst:oapa}}
          \and
          S. Benatti\inst{\ref{inst:oapa}}
          \and
          D. Locci\inst{\ref{inst:oapa}}
          \and
          R. Spinelli\inst{\ref{inst:oapa}}
          \and
          M. Baratella\inst{\ref{inst:eso}}
          \and
          K. Biazzo\inst{\ref{inst:oarm}}
          \and
          J. Maldonado\inst{\ref{inst:oapa}}
          \and
          A. F. Lanza \inst{\ref{inst:OACt}}
          \and
          C. Dorn\inst{\ref{inst:zurich}}
          \and
          P. E. Cubillos\inst{\ref{inst:graz},\ref{inst:OATo}}
          \and
          A. Salmi\inst{\ref{inst:zurich}}
          \and
          A. Maggio\inst{\ref{inst:oapa}}
          \and
          L. Naponiello\inst{\ref{inst:OATo}}
          \and
          F. Marzari \inst{\ref{unipadova}}
          \and
          G. Micela \inst{\ref{inst:oapa}}
          \and 
          V. Fardella \inst{\ref{unipa}, \ref{inst:oapa}}
          \and
          L. Malavolta \inst{\ref{unipadova}, \ref{oapadova}}
          \and
          M. Damasso  \inst{\ref{inst:OATo}}
          \and 
          A. Sozzetti \inst{\ref{inst:OATo}}
          \and
          G. Mantovan \inst{\ref{inst:colombo_padova}, \ref{oapadova}}
          \and 
          D. Nardiello \inst{\ref{unipadova}, \ref{oapadova}}
          \and
          I. Carleo \inst{\ref{inst:OATo}}
          \and 
          R. Claudi \inst{\ref{oapadova},\ref{ist:rm3} }
          \and 
          R. Cosentino\inst{\ref{tng}}
          \and 
          M. Gonzalez\inst{\ref{tng}}
          \and 
          D. Muthukrishna \inst{\ref{MIT}, \ref{CAHarvard}}
          \and
          M. Pinamonti\inst{\ref{inst:OATo}}
          \and 
          T. Zingales\inst{\ref{unipadova},\ref{oapadova}}
          }

   \institute{
    {INAF – Osservatorio Astronomico di Palermo, Piazza del Parlamento, 1, 90134 Palermo, Italy.} \label{inst:oapa}\\ \email{claudia.dimaio@inaf.it}
    \and{ESO - European Southern Observatory, Alonso de Córdova 3107, Casilla 19, Santiago 19001, Chile} \label{inst:eso}
    \and{INAF-Osservatorio Astronomico di Roma, Via Frascati 33, I-00040 Monte Porzio Catone (RM), Italy}\label{inst:oarm} 
    \and{INAF – Osservatorio Astrofisico di Catania, via S. Sofia 78, 95123 Catania, Italy}\label{inst:OACt}
    \and{Institute for Particle Physics and Astrophysics, ETH Zürich, Otto-Stern-Weg 5, 8093 Zürich, Switzerland}\label{inst:zurich} 
    \and{Space Research Institute, Austrian Academy of Sciences, Schmiedlstrasse 6, A-8042, Graz, Austria}\label{inst:graz}
    \and{INAF - Osservatorio Astrofisico di Torino, Via Osservatorio 20, 10025 Pino Torinese, Italy}\label{inst:OATo}
    \and{Dipartimento di Fisica e Astronomia, Università degli Studi di Padova, Vicolo dell’Osservatorio 3, I-35122 Padova, Italy} \label{unipadova}
    \and{Università degli Studi di Palermo, Dipartimento               di Fisica e Chimica, via Archirafi 36, Palermo, Italy.}\label{unipa}
    \and{INAF - Osservatorio Astronomico di Padova, Vicolo dell'Osservatorio 5, I-35122 Padova, Italy} \label{oapadova}
    \and{Centro di Ateneo di Studi e Attivit\`a Spaziali ``G. Colombo'' -- Universit\`a di Padova, Via Venezia 15, IT-35131, Padova, Italy}\label{inst:colombo_padova}
      \and Dipartimento di Matematica e Fisica, Universit\`a Roma Tre, Via della Vasca Navale 84, 00146 Roma, Italy\label{ist:rm3}
    \and{Fundaci{\'o}n Galileo Galilei - INAF, Rambla Jos{\'e} Ana Fernandez P{\'e}rez 7, E-38712 Bre$\tilde{\rm n}$a Baja (TF), Spain}\label{tng}
    \and{Massachusetts Institute of Technology, Cambridge, MA 02139,
USA}\label{MIT}
    \and{Center for Astrophysics, Harvard \& Smithsonian, Cambridge, MA 02138,
USA}\label{CAHarvard}
    }

   \date{Received; accepted }

  \abstract
   {Precise mass and radius measurements of small, transitional exoplanets, such as super-Earths and sub-Neptunes, are essential to constrain their bulk density and formation history, serving as prerequisites for atmospheric characterization.}
   {The ArMS Large Programme, carried out within GAPS using the HARPS-N spectrograph at the Telescopio Nazionale Galileo, aims to confirm and characterize transitional planets in the radius valley through high-precision radial-velocity (RV) measurements. The ultimate goal is to identify ideal targets for atmospheric follow-up observations with next-generation facilities like the James Webb Space Telescope and the future ESA Ariel satellite. We present the first mass determination of a sub-Neptune planet using data entirely collected within the ArMS programme, focusing on the validated planet TOI-4602\,b.}   
   {We monitored TOI-4602, which hosts a close-in validated sub-Neptune (P $\sim$ 3.98 d) detected by the Transiting Exoplanet Survey Satellite (TESS), searching for planet-induced RV variations. We then performed a joint analysis of these RV measurements together with the TESS photometric data. 
   }
   {We determined that TOI-4602\,b is a sub-Neptune with a radius of $R_p = 2.5\pm0.2 \ R_\oplus$ and a mass of $M_p = 5.5\pm0.9 \ M_\oplus$. The resulting bulk density ($\rho_p = 2.1\pm0.6 \ g\ cm^{-3}$) and atmospheric evolution modelling suggest the planet is retaining a tenuous envelope while evolving toward a bare core, consistent with a position immediately above the radius valley.}
   {Given its bright (V = 8.4) and quiet host star and the high Transmission Spectroscopy Metric (TSM) value (140 $\pm$ 54), TOI-4602\,b is a prime target for atmospheric characterization. Simulated retrievals indicate that JWST and Ariel can effectively constrain its atmospheric composition, offering a unique window into the physical processes driving the sub-Neptune to super-Earth transition.}


   \maketitle 
%
\section{Introduction}
The field of exoplanet science recently crossed a major threshold, with the catalogue of confirmed worlds now exceeding 6,000 \citep[NASA Exoplanet Archive,][]{Christiansen2025PSJ.....6..186C} and enabling robust demographic studies of planet populations.
This growing inventory has revealed populations of super-Earths and sub-Neptunes, planet types absent from our Solar System.
Data from the \textit{Kepler} mission  \citep{Borucki2010Sci...327..977B} showed that such small planets ($\rm R_p < 4 R_\oplus$) with short orbital periods ($\rm P_{orb} < 100$ days) are the most common in our Galaxy \citep[see, e.g.,][]{Howard2012ApJS..201...15H, Petigura2013PNAS..11019273P, Fressin2013ApJ...766...81F, Burke2015ApJ...809....8B}. 
The Transiting Exoplanet Survey Satellite \citep[TESS,][]{Ricker2015JATIS...1a4003R} has continued to expand this sample, yet the fundamental nature of these planets remains debated, as their bulk densities can be explained by multiple degenerate compositions \citep[see, e.g.,][]{Rogers2011ApJ...738...59R, Valencia2013ApJ...775...10V}. Although super-Earth and sub-Neptunes overlap in their mass range, they show a clear separation in radius, known as the "radius valley" - an observed paucity of planets around $\rm \sim 1.8 R_{\oplus}$,  whose exact location varies with orbital period and stellar mass \citep[see, e.g.,][]{Owen2013ApJ...775..105O, 2017AJ....154..109F, Petigura2018AJ....155...89P, Petigura2022AJ....163..179P, VanEylen2018MNRAS.479.4786V, Cloutier2020AJ....159..211C, Ho2023MNRAS.519.4056H}. Recent studies have further refined this picture by revisiting the dependence of the valley on stellar irradiation and orbital period, highlighting the role of these parameters in sculpting the transition between rocky and gas-rich worlds \citep[e.g.,][]{Martinez2019ApJ...875...29M, Wanderley2025ApJ...993..233W}. This bimodal distribution, identified by \textit{Kepler}, separates a population of smaller, typically rocky super-Earths from a population of larger sub-Neptunes, whose lower bulk densities suggest the presence of significant volatile components. 

Two main hypotheses have been proposed to explain the origin of this valley. The first is the 'gas dwarf' scenario, where both super-Earths and sub-Neptunes form with solid rock/iron interiors surrounded by thin H$_2$/He-dominated envelopes. Atmospheric escape processes, such as core-powered mass loss \citep[see, e.g.,][]{Ginzburg2016ApJ...825...29G, Ginzburg2018MNRAS.476..759G, Gupta2019MNRAS.487...24G, Gupta2020MNRAS.493..792G} and X-ray and extreme ultraviolet (XUV) photoevaporation \citep[see, e.g.,][]{Lopez2013ApJ...776....2L, Owen2013ApJ...775..105O, Owen2017ApJ...847...29O, Jin2018ApJ...853..163J}, would then strip the envelopes of lower-mass, close-in planets, leaving bare cores and producing the super-Earth population, while more massive planets retain their primordial atmospheres as sub-Neptunes. This evolutionary model is supported by direct observations of the escape of hydrogen and helium \citep[e.g.,][]{DosSantos2023IAUS..370...56D, Guilluy2024A&A...686A..83G}. 
The second hypothesis is the 'water-world' scenario, which posits that the radius valley reflects two distinct formation pathways, producing planets with intrinsically different interior compositions \citep[e.g.,][]{Mordasini2009A&A...501.1139M, Venturini2016A&A...596A..90V, Zeng2019PNAS..116.9723Z, Izidoro2021A&A...650A.152I, Luque2022Sci...377.1211L, Burn2024NatAs...8..463B}. In this case, the smaller-radius population would be rocky, whereas the larger ones would consist of planets with water-rich interiors, formed beyond the water-ice line where volatile condensation enhances the solid reservoir, and later migrated inwards \citep[e.g.,][]{Leger2004Icar..169..499L, Zeng2019PNAS..116.9723Z, Venturini2020A&A...643L...1V}. 
Distinguishing between these scenarios requires detailed atmospheric characterization to detect H$_2$/He envelopes or constrain the abundances of heavier elements, in particular H$_2$O  \citep[e.g.,][]{Seager2000ApJ...537..916S, Fortney2013ApJ...775...80F}. Precise and accurate mass and radius measurements, and hence bulk densities, of sub-Neptunes are then crucial, as they enable robust interior structure modelling \citep[e.g.,][]{Lillo-Box2020A&A...642A.121L, Luque2022Sci...377.1211L, Castro-Gonzalez2023A&A...675A..52C} and provide essential context for atmospheric studies \citep{Batalha2019ApJ...885L..25B, DiMaio2023}. 

In this paper, we report the robust mass measurement of TOI-4602\,b (TIC 467651916), a sub-Neptune discovered by TESS orbiting a G-type star, through the observations of the ArMS Large Programme (LP).  
This paper is organised as follows: in Sect. \ref{sec:ArMS} we describe the ArMS LP. We present the photometric and radial velocity observations in Sect. \ref{sec:dataset}, and the stellar characterization in Sect. \ref{sec:hoststar}. The analysis of the planetary signal is presented in Sect. \ref{sec:analysis}. In Sect. \ref{sec:internal} we discuss about TOI-4602\,b composition and investigate its atmospheric evolution in Sect. \ref{sec:fotoevaporazione}. We evaluate the potential for atmospheric retrieval with Ariel in Sect. \ref{sec:ariel_retrieval} and draw our conclusions in Sect. \ref{sec:discussion_conclusions}.


\section{The ArMS Large Programme}\label{sec:ArMS}
The Ariel Masses Survey (ArMS, PI: S. Benatti, see also \citealt{Bonomo2026A&A...707A.197B}) is a 5-year LP started in October 2023 exploiting the HARPS-N spectrograph \citep{2014SPIE.9147E..8CC} at the Telescopio Nazionale Galileo (TNG) in La Palma (Canary Islands). 
The main goal of the project is to investigate the composition of transitional planets between the super-Earth and sub-Neptune families, falling within the Radius Valley. 
We aim at providing robust mass determinations for small transiting planets suitable for atmospheric characterisation, e.g. by the James Webb Space Telescope (JWST, \citealt{2006SSRv..123..485G}) and the future ESA M4 Ariel mission \citep{2018ExA....46..135T}. Since bulk density alone yields degenerate composition models (e.g. \citealt{Zeng2019PNAS..116.9723Z}), obtaining a mass precision of at least 30\% is crucial for reliable atmospheric retrieval and to unveil the true nature of these objects \citep{DiMaio2023}.

ArMS selects targets primarily from the continuously updated Ariel candidate lists, focusing on planets with radii between 1.3 and 2.5 R$_{\oplus}$ and poorly constrained masses. A subsample targets young stars (< few hundred Myr) to investigate star-planet interactions and photo-evaporation (e.g. \citealt{Locci2024PSJ.....5...58L}), extending the radius selection to 3.5 R$_{\oplus}$ for these systems.

ArMS operates within the GAPS3 collaboration, leveraging expertise and data from previous GAPS programmes (see, e.g. \citealt{2013A&A...554A..28C,2020A&A...638A...5C}). In this paper, we present the first mass determination of a sub-Neptune derived entirely from ArMS observations.

\section{Description of the dataset}\label{sec:dataset}
\subsection{TESS photometric observations}
\begin{figure}[h!]
	\centering
	\includegraphics[scale=0.5]{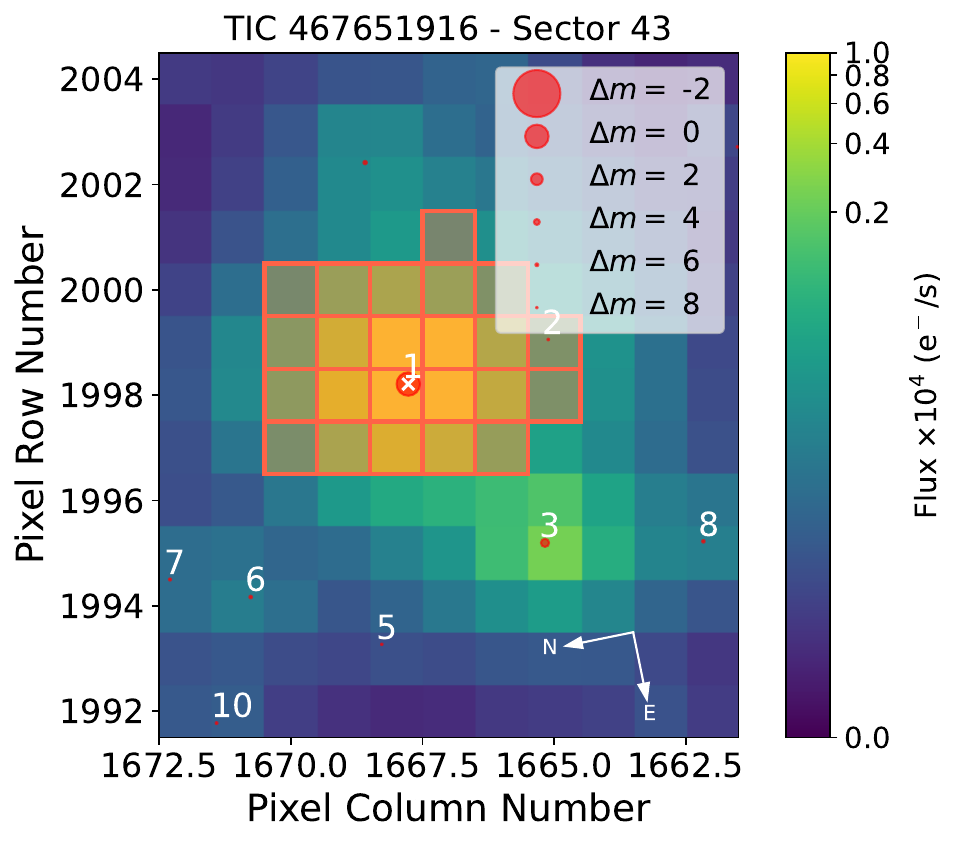}
	\caption{Target pixel file from the TESS observations in Sector 43, centred on TOI-4602, which is highlighted with a white cross. The orange squares indicate the SPOC pipeline aperture, while sources from \textit{Gaia} DR3 catalogue \citep{GaiaCollaboration2023A&A...674A...1G} are overplotted with circles whose sizes are proportional to the  \textit{Gaia} magnitude difference with respect to our target. Gray arrows show the direction of the proper motions for all sources in the field. The colour bar indicates the electron counts per pixel.}
	\label{fig:TPF}
\end{figure}

\begin{figure*}[h!]
	\centering
	\includegraphics[scale=0.5]{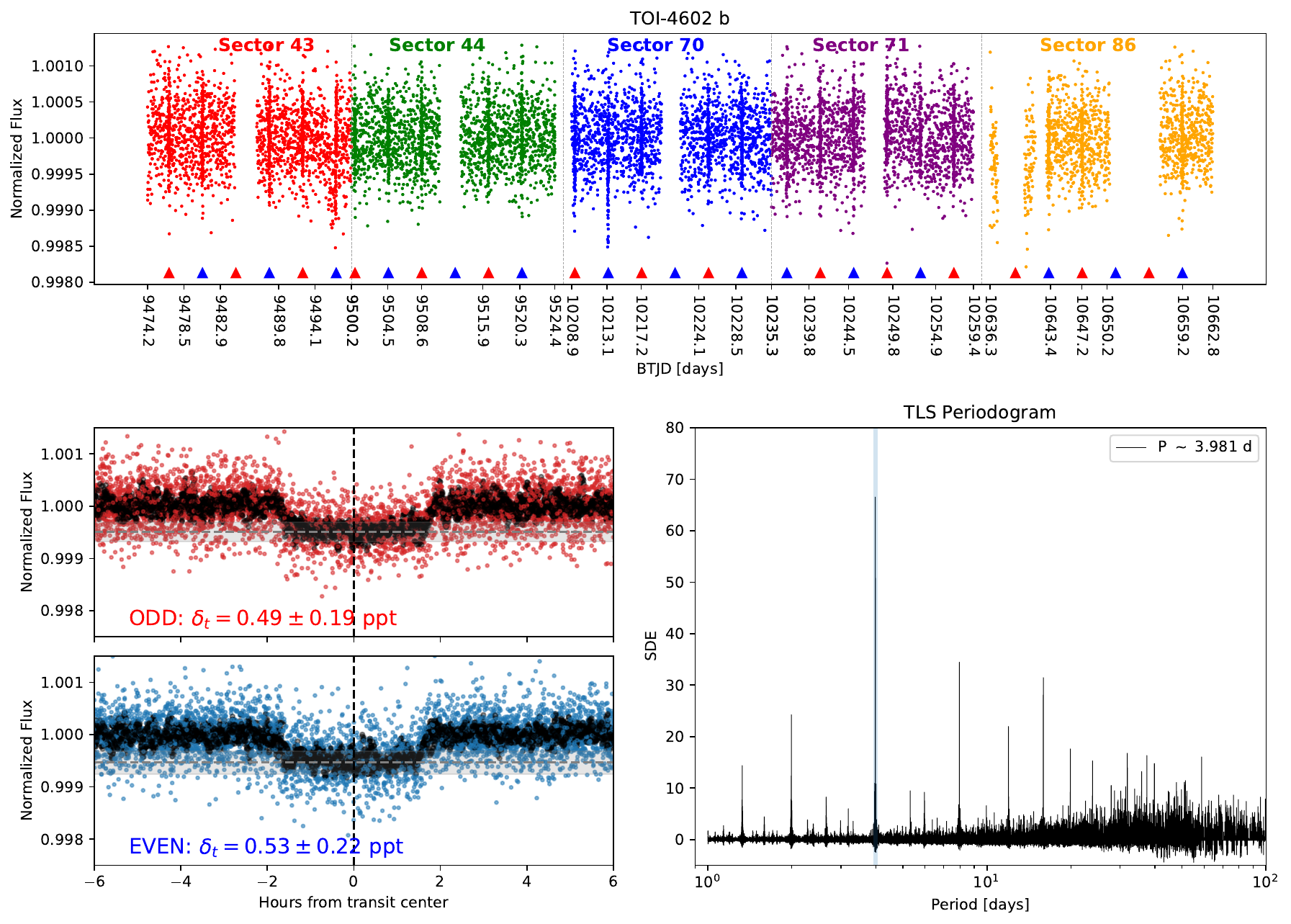}
	\caption{(Top panel) Normalized TESS PDCSAP light curves from all available sectors. The out-of-transit data have been binned to 20 minutes to improve the signal-to-noise ratio, while the in-transit data are kept at the original cadence to preserve the transit shape. Red and blue triangles mark the positions of odd and even transits, respectively. BTJD refers to the Baycentric TESS Julian Date, which is the time stamp measured in BJD and offset by 2\,450\,000.0. Dots in different colours represent different TESS sectors, with vertical dashed lines indicating the boundaries between them. (Bottom left panels) Odd and even transits folded using a period of 3.98 days. The consistent transit depths between odd and even transits show no evidence of alternating primary and secondary eclipses, disfavouring an eclipsing binary scenario at twice the orbital period. Black curves show smoothed light curves obtained via a moving average. (Bottom right panel) TLS periodogram of TOI 4602\,b light curves. The highest peak corresponds to the main transit signal at P $ \approx 3.98$ days. The additional secondary peaks visible in the periodogram correspond to harmonics (e.g., P$/2$, $2$P) and aliases of this primary periodicity.}
		\label{fig:sectors}
	\end{figure*}

TOI-4602 was observed photometrically by TESS in short-cadence mode (120 s) across multiple sectors: Sector 43 (UT 2021 September 16 - October 11, GO4191), Sector 44 (UT 2021 October 12 - November 05, GO4191), Sector 70 (UT 2023 September 21 - October 16, program IDs: GO6200, GO6032, GO6144), Sector 71 (UT 2023 October 17 - November 11, program IDs: GO6200, GO6032, GO6144), and Sector 86 (UT 2024 November 21 - December 18). In this work, we used the light curves extracted from all these datasets. We adopted the Pre-search Data Conditioning Simple Aperture Photometry (PDCSAP) flux, corrected for time-correlated instrumental systematics, as provided by the TESS SPOC pipeline \citep{Jenkins2016SPIE.9913E..3EJ}. These data were retrieved using the Python package \texttt{lightkurve} \citep{Lightkurve2018} from the Mikulski Archive for Space Telescopes. 
During preprocessing, we excluded all data point with  \texttt{DQUALITY}$>$0 to ensure the highest data quality for our analysis. To verify the absence of contaminating sources within the TESS aperture, we inspected the Target Pixel Files (TPF), which contain the original CCD pixel observations, using \texttt{tpfplotter}
\citep{Aller2020A&A...635A.128A}. This tool overlays all sources from the \textit{Gaia} Data Release 3 within a specific contrast magnitude relative to our target (here, we adopted $\Delta m = 8$) and highlights the aperture mask employed by the SPOC pipeline. 
To assess the impact of nearby stars on the TESS photometry, we performed a quantitative contamination analysis following the methodology outlined in \citet{Mantovan2022MNRAS.516.4432M}. We used Gaia DR3 data to identify all sources in the vicinity of TOI-4602 and calculated the dilution factor, defined as the ratio of the total contaminating flux within the aperture to the flux of the target star. We identified two primary sources that could contribute to the flux: Gaia DR3 169255938060729984 ($\Delta G \approx 7.3$, separation $\approx 58''$) and Gaia DR3 169255766262035584 ($\Delta G \approx 2.8$, separation $\approx 82''$), labelled as target 2 and 3 in Fig. \ref{fig:TPF}. Although both sources lie near the edge or outside of the SPOC aperture, we calculated their individual contributions to the total contamination. We found their flux ratios to be $0.00081$ and $0.00296$, respectively. The resulting total dilution factor is $\sim 0.004$ (0.4\%). Given that this value is extremely low and its impact on the planetary radius measurement is smaller than the formal uncertainties, we did not apply a dilution correction to the light curves. The normalized TESS PDCSAP light curves from all available sectors are shown in Fig. \ref{fig:sectors}. 

\subsection{HARPS-N spectroscopic observations}
\begin{figure*}[h!]
	\centering
	\includegraphics[scale=0.45]{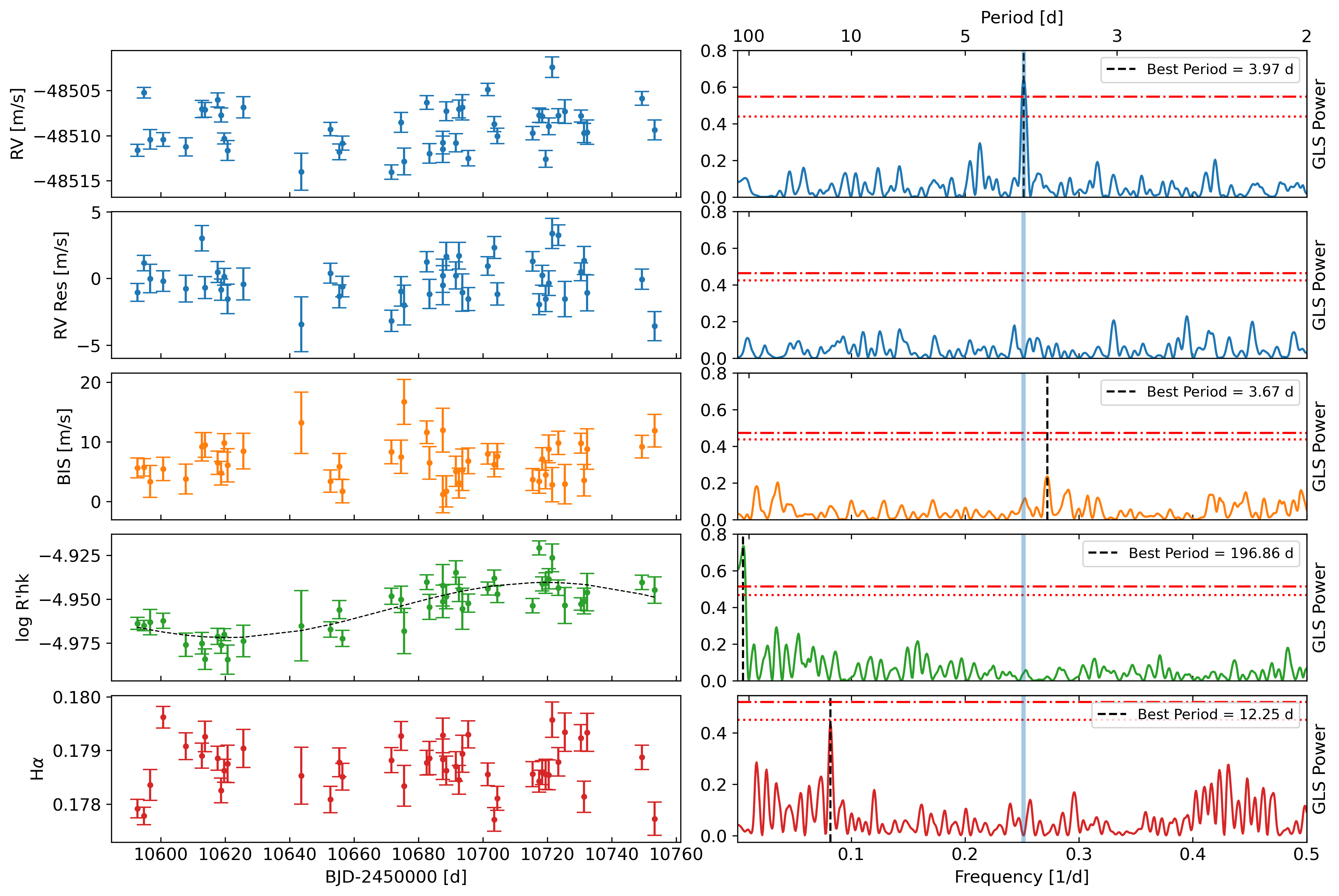}
	\caption{Spectroscopic time series and GLS periodogram. Left: HARPS-N spectroscopic time series analysed in this work. From top to bottom, the panels show the RVs, RV residuals, BIS, $\rm \log R'_{HK}$, and H$\alpha$ measurements. In the $\rm \log R'_{HK}$ panel, a sinusoidal modulation - plotted as a black dashed line - is visible, suggesting the presence of a stellar activity cycle, consistent with the period identified in the corresponding GLS periodogram.  Right: GLS periodograms of the RVs and the spectroscopic activity indicators. The primary peak in each periodogram is marked with a vertical black dashed line. The vertical blue shaded area highlights the period signal at 3.981 days, consistent with the transit-like periodicity identified in the TLS analysis. The horizontal dash-dotted and dotted lines mark, respectively, the 0.1\% and the 1\% FAP levels as evaluated using the bootstrap method. }
		\label{fig:timeseries}
	\end{figure*}

TOI-4602 is among the targets included in the ArMS Large Programme, monitored with HARPS-N at TNG. Observations of this star were conducted between 9 October 2024 and 18 March 2025, resulting in a total of 45 spectra with an average signal-to-noise ratio (S/N) of $\sim 129$ and a median exposure length of 900 s. The HARPS-N spectra were reduced using the standard Data Reduction Software (DRS) pipeline \citep[][and references therein]{Pepe2002A&A...388..632P}, executed through the YABI workflow interface hosted at the IA2 Data Center \footnote{\url{https://www.ia2.inaf.it}}
. The radial velocity (RV) values, reported in Table \ref{tab:timeseries}, were extracted from the Cross-Correlation Function (CCF), computed using a G2 spectral mask with a half-window width of 20 km s$^{-1}$. We achieved RV measurements with a dispersion of few m s$^{-1}$ ($\sim 2.6$ m s$^{-1}$) and an internal error of approximately 1 m s$^{-1}$. In addition, the HARPS-N DRS provides activity diagnostics derived from the CCF, such as the CCF bisector span (BIS) and the full width at half-maximum (FWHM) of the CCF. The log $R'_{HK}$ index from the Ca II H\&K lines was computed using YABI workflow implementation based on the prescriptions of \citet{Lovis2011arXiv1107.5325L} and reference therein, adopting the $(B-V)$ colour index listed in Table \ref{tab:stellar_parameters}. We also extracted the H$_\alpha$ activity index using the ACTIN 2 code \citep{GomesdaSilva2018,GomesdaSilva2021A&A...646A..77G}. In the following analysis, we excluded the spectrum acquired at epoch BJD 2460595.7562 due to its low signal-to-noise ratio (S/N<50) and its RV uncertainty, which lies above the 99.87th percentile of the RV uncertainty distribution. The spectroscopic time series are presented in Figure \ref{fig:timeseries} and detailed in Table \ref{tab:timeseries}.   

\section{The host star}\label{sec:hoststar}

\subsection{Spectroscopic stellar properties}
\label{sec:spec_star_prop}
To derive the effective temperature ($T_{\mathrm{eff}}$), the surface gravity ($\log g$), the micro-turbulence velocity parameter ($\xi$) and the iron abundance ($\rm [Fe/H]$) of the host star, we analysed the co-added spectrum built from individual HARPS-N spectra extracted with the standard DRS pipeline. The final signal-to-noise ratio of the co-added spectrum was $\sim 770$ at $\sim 6000 \ \AA$. 
Given the relatively old age of the target, we applied the standard spectroscopic analysis method based on equivalent widths (EW) of iron lines to derive the stellar parameters. 
Specifically, $\rm T_{eff}$ and $\rm \log g$ were inferred by imposing excitation and ionization equilibria in LTE, while $\xi$ was determined by minimizing the trend between $\rm Fe\,I$ abundances and the reduced equivalent widths $\log(\rm{EW}/\lambda)$.

The EWs of carefully selected iron (Fe) lines (see \citealt{2020baratella} for the complete line list) were measured using the automatic tool ARES (v2.0, \citealt{2015sousa}). Lines with uncertainties larger than 10$\%$ were rejected from the analysis, together with strong lines with EW$> 120$ m\AA.  We finally used the \textit{pymoogi} code \citep{2017adamow}, which is a Python wrapper of the MOOG code \citep{Sneden1973ApJ...184..839S}, to derive the stellar parameters and the Kurucz model atmospheres linearly interpolated from the grid by \cite{2003castelli}. The final values are reported in Table \ref{tab:stellar_parameters}. Our inferred spectroscopic $T_{\mathrm{eff}}$ agrees perfectly with the estimate obtained using various photometric colour indexes and the \textit{colte} tool developed by \citealt{2021casagrande}. The photometric temperatures indeed vary between 5825$\pm$117\,K in $(R_P-J)$ and 6027$\pm$77\,K in $(G-B_P)$, with a weighted mean of 5946$\pm$58\,K. The spectroscopic $\log g$ of 4.34$\pm$0.10 dex is consistent with the value derived from $Gaia$ parallax and the mass value of 1.11 M$_{\odot}$ reported in \cite{tic}.
We also derived the projected rotational velocity ($v\sin i$) of TOI-4602 using the calibration between the FWHM of the CCF and the $v\sin i$ established by \cite{raineretal2023}. The resulting value is presented in Table\,\ref{tab:stellar_parameters}.
We also detected the presence of the lithium absorption line at $\sim$6707.8\,\AA. We measured an equivalent width of $EW_{\rm Li}=36.5\pm0.5$\,m\AA\,and, adopting the spectroscopic parameters measured in this work, we determined a lithium abundance of $A({\rm Li})_{\rm NLTE}$=2.26$\pm$0.04\,dex, after applying the corrections for non-local thermodynamic equilibrium effects by \cite{Lindetal2009}. 
Based on empirical isochrones by \cite{Jeffriesetal2023}, and using the $EW_{\rm Li}$ and $T_{\rm eff}$ values with the {\sc eagles} code, we derive an age of $3.2^{+4.1}_{-2.4}$\,Gyr. The value of the lithium abundance, combined with the effective temperature derived above, places the target coherently with members of the M67 open cluster ($\sim 4.5$\,Gyr; \citealt{pasquinietal2008}). 

\subsection{Mass, radius and age}\label{sec:mass_radius_age}
The Galactic spatial-velocity components $(U, V, W)$ of TOI-4602 are computed from the stellar RV together with {\it Gaia} parallaxes and proper motions \citep{GaiaCollaboration2023A&A...674A...1G} following the procedure described in \citet{Maldonado2010}. To correct these velocities to the Local Standard of Rest (LSR), we adopted the recent solar motion parameters $(U_\odot, V_\odot, W_\odot) = (12.9, 12.0, 7.8)$ km s$^{-1}$ from \citet{Bienayme2024A&A...689A.280B}. The derived values are provided in Table~\ref{tab:stellar_parameters}.
To classify the star as belonging to the thin/thick disc population, we adopted the methodology developed by \citet{Bensby2003,Bensby2005}, based on the relative kinematic probabilities $P_{\text{thick}}/P_{\text{thin}}$ and $P_{\text{thick}}/P_{\text{halo}}$. We derived the following values: $P_{\text{thin}} = 2.13 \times 10^{-8}$, $P_{\text{thick}} = 1.56 \times 10^{-8}$, and $P_{\text{halo}} = 4.99 \times 10^{-11}$. These result in a relative probability ratio of $P_{\text{thick}}/P_{\text{thin}} = 0.73$, with $P_{\text{thick}}/P_{\text{halo}} \gg 1$. According to the adopted framework, thin disk stars are defined by $P_{\text{thick}}/P_{\text{thin}} < 0.1$, while thick disk stars require $P_{\text{thick}}/P_{\text{thin}} > 10$. Since our value falls within the intermediate range ($0.1 < P_{\text{thick}}/P_{\text{thin}} < 10$), TOI-4602 is formally classified as a transition star showing intermediate kinematics.
Following the method of \cite{Almeida-Fernandes2018MNRAS.476..184A}, we also estimated a kinematic age of $11 \pm 3$ Gyr, which, despite the inherent uncertainties of this method for single objects, points toward an old stellar population.

Moreover, we also computed stellar ages (and masses) from isochrone fittings (and evolutionary tracks). We therefore considered the \texttt{PARSEC} model by \citet{Bressan2012MNRAS.427..127B} and the \texttt{PARAM} interface\footnote{\url{https://stev.oapd.inaf.it/cgi-bin/param}}  (version 1.5,  \citealp{daSilva2006A&A...458..609D}).  This code considers as input some observational parameters (effective temperature, parallax, apparent V magnitude, and iron abundance) to perform a Bayesian determination of the most likely stellar intrinsic properties, appropriately weighting all the isochrone sections that are compatible with the observational parameters.
A flat distribution of ages with a range of 0.1–15 Gyr was considered as priors for the analysis. We considered as effective temperature, stellar gravity and iron abundance those values we derived as described in Section \ref{sec:spec_star_prop}. 
We obtained in this way a system age of 8.7$^{+1.3}_{-1.2}$ Gyr , a stellar mass of 0.90$\pm$0.02 $\rm M_\odot$ and a stellar radius of R = 1.08$^{+0.10}_{-0.07} \rm R_\odot$, where the uncertainties are those provided by the \texttt{PARAM} tool and do not include possible systematic uncertainties in the adopted stellar models.

\begin{table} \footnotesize										
\centering										
\caption{Stellar parameters of TOI-4602.}										
\label{tab:stellar_parameters}
\begin{tabular}{lcc}										
\hline										
\hline										
\rule{0pt}{2.8ex}Parameter (Unit) & Value & Source \\			
\hline										
\multicolumn{3}{l}{\normalsize \rule{0pt}{2.6ex}Identifiers} \\[0.4ex]											
TOI			&	4602	&	TOI catalog	\\			
TIC			&	467\,651\,916	&	TIC	\\			
HIP			&	18\,841	&	HIP	\\	
HD          &   25295   &   HD \\
										
TYC			&	2357-595-1	&	Tycho	\\			
2MASS			&	J04022420+3119501	&	2MASS	\\			
\textit{Gaia}			&	169256041140206848	&	\textit{Gaia} DR3	\\			
										
\multicolumn{3}{l}{\normalsize \rule{0pt}{2.6ex}Astrometric properties} \\[0.4ex]										
$\alpha$ (J2000.0)			&	04:02:24.219	&	\textit{Gaia}	DR3	\\		
$\delta$ (J2000.0)			&	+31:19:50.39	&	\textit{Gaia}	DR3	\\		
$\pi$ (mas)			&	15.921	$\pm$	0.028	&	\textit{Gaia} DR3	\\	
$\mu_\alpha\cos\delta$ (mas yr$^{-1}$)			&	-115.397	$\pm$	0.033	&	\textit{Gaia} DR3	\\	
$\mu_\delta$ (mas yr$^{-1}$)			&	-192.412	$\pm$	0.023	&	\textit{Gaia} DR3	\\	
$Distance$ (pc)			&	62.64 $\pm$ 0.11			&	\textit{Gaia} DR3	\\	
										
\multicolumn{3}{l}{\normalsize \rule{0pt}{2.6ex}Photometric properties} \\[0.4ex]										
$B_T$			&	9.028	$\pm$	0.012	&	Tycho-2	\\	
$V_T$			&	8.392	$\pm$	0.010	&	Tycho-2	\\	
$(B-V)_0$       & 0.591 & Tycho-2	\\	
$G_{\rm BP}$			&	8.4709	$\pm$	0.0028	&	\textit{Gaia}	DR3	\\
$G$			&	8.1839	$\pm$	0.0028	&	\textit{Gaia}	DR3	\\
$G_{\rm RP}$			&	7.7227	$\pm$	0.0038	&	\textit{Gaia}	DR3	\\
$J$			&	7.196	$\pm$	0.020	&	2MASS	\\	
$H$			&	6.934	$\pm$	0.031	&	2MASS	\\	
$K_S$			&	6.883	$\pm$	0.027	&	2MASS	\\	
$W1$			&	6.868	$\pm$	0.070	&	AllWISE	\\	
$W2$			&	6.859	$\pm$	0.020	&	AllWISE	\\	
$W3$			&	6.904	$\pm$	0.017	&	AllWISE	\\	
$W4$			&	6.878	$\pm$	0.082	&	AllWISE	\\	
$A_V$			&	$<$0.12	&	This work	\\			
										
\multicolumn{3}{l}{\normalsize \rule{0pt}{2.6ex}Stellar parameters} \\[0.4ex]										
Spectral type			&	G5		&	Tycho-2	\\		
L$_\star$ (L$_\odot$)			&	1.3$^{+0.3}_{-0.2}$	&	This work	\\	
M$_\star$ ($M_\odot$)			&	0.90	$\pm$	0.02	&	This work	\\	
R$_\star$ ($R_\odot$)			&	1.08$^{+0.10}_{-0.07}$			&	This work	\\	
$T_{\rm eff}$ (K)			&	5966	$\pm$	50	&	This work	\\	
$v\sin i$ (km s$^{-1}$)			&		2.2$\pm$0.7		&	This work	\\	
$\log$ g$_\star$ [cgs]			&	4.34	$\pm$	0.10	&	This work	\\	
$\xi$ (km s$^{-1}$)			&		1.25 $\pm$	0.08	&	This work	\\	
{[Fe/H]} [dex]			&	-0.26	$\pm$	0.03	&	This work	\\	
$A({\rm Li})_{\rm NLTE}$            &     2.26$\pm$0.04  &  This paper   \\
$\rho$$_\star$ ($\rho$$_\odot$)			&	0.71$^{+0.15}_{-0.16}$			&	from PARAM	\\	
$\log R'_{\rm HK}$ [dex]		&	-4.955	$\pm$	0.015	&	This work	\\	
Age (Gyr)			&	3.2$^{+4.1}_{-2.4}$			&	from EW$_{Li}$	\\	
Age (Gyr)			&	$\sim$ 4.5			&	from logNLi$_{NLTE}$	\\	
Age (Gyr)			&	8.7$^{+1.3}_{-1.2}$			&	from PARAM	\\	
$U^{(a)}$ (km s$^{-1}$)			&	 58.15 $\pm$ 0.02 &	This work	\\			
$V^{(a)}$ (km s$^{-1}$)			&	-31.41 $\pm$ 0.04 &	This work	\\			
$W^{(a)}$ (km s$^{-1}$)			&	-49.48 $\pm$ 0.11 &	This work	\\						
										
\hline										
\end{tabular}										
										
\vspace{0.5em}										
\raggedright										
\textbf{Notes.} TESS Primary Mission TOI catalogue \citep{Guerrero2021ApJS..254...39G}; TIC \citep{Stassun2018AJ....156..102S, Stassun2019AJ....158..138S}; HIP \citep{Perryman1997A&A...323L..49P}; Tycho-2 \citep{Hog2000A&A...355L..27H}; 2MASS \citep{Skrutskie2006AJ....131.1163S}; \textit{Gaia} DR3 \citep{GaiaCollaboration2023A&A...674A...1G}; AllWISE \citep{Cutri2014yCat.2328....0C}. \\										
(a) Galactic velocity components, where $U$ is positive towards the Galactic centre, $V$ in the direction of the Galactic rotation, and $W$ towards the North Galactic Pole.										
\end{table}

\subsection{Stellar activity}\label{sec:activity}

We characterized the activity level of the host star by analysing the chromospheric log R'$_{\rm HK}$ index. 

The time series (see Fig.\ref{fig:timeseries} and Table \ref{tab:timeseries}) shows a stable mean $\log R'_{HK} \approx -4.95$ using standard (B-V) calibration. Adopting the $T_{\text{eff}}$-based calibration by \citet{Lorenzo-Oliveira2018A&A...619A..73L}, we find $\log R'_{HK} \sim -5.05$. Using the recent relation by \cite{Carvalho-Silva+2025} for [Fe/H] = -0.3, this yields an age of 8.0-9.5 Gyr. This range is consistent with both the isochronal 
and kinematic ages
, indicating TOI-4602 is an old star likely in the 8.0–9.0 Gyr range.

Given the low activity level, the search for a rotational signal proved to be challenging. We analysed both TESS photometry and spectroscopic activity indicators to determine the stellar rotation period (P$_{\mathrm{rot}}$), but we did not obtain a direct measurement. A Generalized Lomb-Scargle (GLS, \citealt{Zechmeister2009A&A...496..577Z}) periodogram analysis revealed no significant periodicity in either data, although a minor, statistically not significant peak is visible in the H$\alpha$ periodogram. The only noteworthy signal is a long-period peak in the log R'$_{\rm HK}$ periodogram, which we interpret as a likely magnetic activity cycle rather than rotation. 
Using the mean $\rm \log R'_{\rm HK}$ value and the star's B-V colour (Table \ref{tab:stellar_parameters}), the relation from \citet{Noyes1984ApJ...279..763N} yields an empirical period of P$_{\mathrm{rot}} \sim$ 17 days. 

Finally, in the absence of direct X-ray observations of TOI-4602, we adopted the prediction of a X-ray luminosity between $L_{\rm x} = 6 \times 10^{26}$ and $2 \times 10^{27}$\,erg s$^{-1}$ based on the stellar age \citep{Mamajek+Hillenbrand2008}. Hence,
we assume $L_{\rm x} = 1^{+1.0}_{-0.4} \times 10^{27}$\,erg s$^{-1}$ for deriving the high-energy irradiation of the planetary atmosphere, in line with the expectation for a Sun-like star.

\section{Analysis}\label{sec:analysis}

\subsection{Detection and vetting of TOI-4602\,b}
An alert for a planet candidate around TOI-4602 was released by the TESS Mission on 2021 November 2, when the SPOC pipeline \citep{Jenkins2016SPIE.9913E..3EJ} at NASA Ames Research Center identified a candidate exoplanet with a period of 3.98 days. The planetary nature of this target has already been statistically validated using \texttt{TRICERATOPS} \citep{Giacalone2020ascl.soft02004G} by \citet{Hord2024AJ....167..233H}. In this section, we provide additional evidence supporting the planetary nature of TOI-4602\,b. 
We first detrended (flattened) the light curves of TOI-4602 using the Savitzky-Golay filter implemented in \texttt{lightkurve}, masking the likely transit events based on the ephemeris provided by the SPOC pipeline to build the model. We then computed the Transit Least Squares (TLS) periodogram of the detrended light curve, which combines all available TESS sectors, to search for transit signals with periods between 1 and 100 days using the Python package \texttt{transitleastsquares} \citep{Hippke2019A&A...623A..39H} . The periodogram reveals a significant peak at $P \sim 3.98$ days, with a Signal Detection Efficienty (SDE) of $\sim 45$, and a stacked S/N $\sim 9$. In Fig. \ref{fig:sectors}, the positions of the odd and even transits are indicated with red and blue triangles, respectively. To verify the consistency of the transit depths, we compared the phase-folded light curves of odd and even transits. The resulting difference in depth corresponds to a significance of approximately 0.1$\sigma$, indicating that the depths of odd and even transits are consistent within the uncertainties and supporting the planetary nature of TOI-4602\,b. 

\subsection{Photometry time-series analysis}\label{sec:photometry_analysis}
To assess the transit signal robustness, we computed TLS periodograms for individual TESS sectors, confirming the $\sim$3.98-day signal. 
An inspection of the residuals revealed no other significant peaks, suggesting no evidence of additional transiting planets within the period range probed by TESS.

We performed a global photometric modelling of all TESS sectors within a Bayesian framework using \texttt{PyORBIT} \citep{Malavolta2016A&A...588A.118M,Malavolta2018AJ....155..107M}, which is adopted consistently throughout this work for all orbital analysis. We simultaneously modelled each transit using the \texttt{batman} package \citep{Kreidberg2015PASP..127.1161K}, fitting the following parameters: orbital planetary period (P), central time of transit ($T_0$), planetary-to-star ratio ($R_p/R_\star$), impact parameter $b$, stellar density ($\rho_\star$, in solar units), quadratic limb-darkening (LD) coefficients $u_1$ and $u_2$ adopting the LD parametrization ($q_1$ and $q_2$) introduced by  \citet{Kipping2013MNRAS.435.2152K}, and photometric jitter terms to account for unmodelled effects (e.g., short-term stellar activity) or any underestimated uncertainties. 
Before proceeding with the \texttt{PyORBIT} analysis, we estimated the limb-darking coefficients and their uncertainties assuming a quadratic law using the \texttt{PyLDTk} package \citep{Parviainen2015MNRAS.453.3821P}, which is based on the spectrum library by \citet{Husser2013},  by computing them specifically for the TESS passband. Using the effective temperature, log \textit{g}, and metallicity reported in Table \ref{tab:stellar_parameters}, we initially derived $u_1 = 0.3876 \pm 0.0003$ and $u_2 = 0.1413 \pm 0.0008$. Following \citet{Patel2022AJ....163..228P}, we adopted broader Gaussian priors with uncertainties of 0.2 for both coefficients to mitigate potential systematic discrepancies between theoretical models and TESS observations during the \texttt{PyORBIT} global fit.
In addition, we adopted Gaussian priors of the following parameters (see Table \ref{tab:join_fit_results} for details): the orbital period, as indicated by the TLS periodogram; the transit time given by the first predicted transit, and the stellar density ($0.71 \pm 0.16\,\rho_\odot$) estimated from the stellar mass and radius in Table \ref{tab:stellar_parameters}. 
Following standard procedure with \texttt{PyORBIT}, we first performed a global optimization of the model parameters by executing a differential evolution algorithm implemented via \texttt{PyDE}\footnote{\url{https://github.com/hpparvi/PyDE}} \citep{Storn1997JGOpt..11..341S}. This was followed by a Bayesian analysis of each selected light curve around each transit using the affine-invariant ensemble sampler \citep{Goodman2010CAMCS...5...65G} for Markov chain Monte Carlo (MCMC) implemented within the package \texttt{emcee} \citep{Foreman-Mackey2013PASP..125..306F}. We employed 8$n_{dim}$ walkers (where $n_\mathrm{dim}$ is the number of free parameters) for 50\,000 generations with \texttt{PyDE}, followed by 150\,000 steps with \texttt{emcee}, applying a thinning factor of 100 to reduce the effects of the chain auto-correlation. The first 25\,000 steps were discarded as burn-in phase, after verifying convergence through the Gelman-Rubin statistic \citep{Gelman1992StaSc...7..457G}. According to this analysis, the planetary radius is \rm $R_p$ = 2.45$\pm$0.23 R$_\oplus$ with TOI-4602\,b orbiting around its star with a planetary period of 3.9812999$^{+0.0000049}_{-0.0000046}$ days, which is compatible with the results obtained from the TLS analysis and those given by the SPOC pipeline.  

To search for perturbations potentially induced by additional non-transiting planets, or by bodies not detected by TESS, we investigated the presence of transit timing variations (TTVs) using the approach described in \citet{Naponiello2026A&A...705A...5N}. 
We don't find compelling evidence for significant timing modulations with a semi-amplitude upper limit of 3.5 minutes. In particular, the difference in BIC between the linear and sinusoidal models is slightly below the commonly adopted significance threshold of 10 (Fig.\,\ref{fig:TTV}), and the scatter statistic, $\chi^2_{\mathrm{mod}}$ \citep{Naponiello2026A&A...705A...5N}, is close to unity, supporting this non-detection.

\subsection{Spectroscopic time-series analysis}
Figure \ref{fig:timeseries} shows the time series and the GLS periodograms of HARPS-N RVs and all activity indicators. Regarding the planetary signals, we identified a strong peak in the RV periodogram (FAP $<$ 0.1\%) at the same period as the photometric signal. This feature is absent in all stellar activity indicators, supporting a planetary origin for the signal. To verify the presence of additional companions, we subtracted the sinusoidal model corresponding to this main peak from the RV time series (pre-whitening). The resulting residuals and their corresponding GLS periodogram, shown in the second panels from the top in Figure \ref{fig:timeseries}, do not exhibit any further significant peaks, indicating that no other signals are detectable above the noise level. In terms of stellar activity, a long-term trend is visible in the $\rm \log R'_{HK}$ time series, highlighted by the sinusoidal fit overplotted in the panel. The corresponding periodogram shows a significant peak around 197 days, which could suggest a stellar magnetic cycle. However, as our baseline covers less than one complete cycle, this period remains approximate. Aside from this long-term modulation, no other significant peaks were detected in the activity periodograms, consistent with the average value of the $\rm \log R'_{HK} = -4.955 \pm 0.015$, which indicates a low level of stellar activity for the host star. 

Following the detection of a significant RV signal, we proceeded with the full orbital analysis of the RV dataset of TOI-4602 using \texttt{PyORBIT}, with the aim of determining the planetary mass. We adopted the same \texttt{PyORBIT} configuration as described in Section \ref{sec:photometry_analysis}. In addition to the previous parameters, we included the systemic RV (offset), and the RV semi-amplitude (K) as free parameters.   
We tested the following models: 
\begin{itemize}
    \item[i.] A single planet on a circular orbit, without stellar activity correction (Model 1, M1).
    \item[ii.] Two planets on circular orbits, to investigate the possibility of an additional companion (Model 2, M2).
    \item[iii.] A single planet on a circular orbit with quasi-periodic Gaussian Process (GP) to account for potential stellar activity signals (Model 3, M3). 
    \item[iv.] A single planet on an eccentric orbit (Model 4, M4).
\end{itemize}
We obtained the following Bayesian Information Criterion (BIC) values: BIC(M1) = 188.8, BIC(M2) = 199.5, BIC(M3) = 204.9, BIC(M4) = 196.0. To select the best model, we follow the criteria from \citet{Kass01061995}, where evidence favouring the model with the lower BIC is considered 'strong' for $|\Delta$BIC|$>$ 6 and 'very strong' for $|\Delta$BIC|$>$ 10. Based on this, we find that M1 is very strongly preferred over M2 and M3 ($\Delta$BIC$_{21}$ = 10.7, $\Delta$BIC$_{31}$ = 16.1), and strongly preferred over M4 ($\Delta$BIC$_{41}$ = 7.2).

In addition to these scenarios, we explicitly tested the inclusion of a linear velocity trend to account for potential long-term drifts. This model yielded a $\text{BIC} = 192.47$, resulting in a $\Delta\text{BIC} = 3.67$ in favour of the simpler 1-planet circular model (M1). We also explored a broader two-planet scenario by searching for additional Keplerian signals with periods up to 300 days, which was decisively disfavoured with a $\text{BIC} = 207.23$ ($\Delta\text{BIC} \approx 18.4$ relative to M1). Considering the results of all these tests, we find that M1 is statistically preferred over all other configurations.
Consequently, we adopt Model 1 as the best representation of the RV data, from which we derived a RV semi-amplitude of $2.59^{+0.34}_{-0.35}$ m/s, a planetary mass of $5.98^{+0.80}_{-0.82} M_{\oplus}$ for TOI-4602\,b, and an orbital period of $3.981^{+0.010}_{-0.011}$ days, in excellent agreement with the photometric period.
We further quantify the detection sensitivity of our RV time-series in Appendix\,\ref{sec:completeness}, along with the Proper Motion anomaly (PMa) revealed for this star by the Hipparcos-Gaia Catalogue of Accelerations (HCGA; \citealt{Brandt2018}).

\subsection{Joint radial velocity and photometry analysis}

A joint transit and RV analysis was performed using \texttt{PyORBIT}. 
For this global modelling, we switched from the MCMC sampler used in the previous sections to the Multinest Nested Sampling algorithm via \texttt{PyMultiNest} \citep{Feroz2009MNRAS.398.1601F, Buchner2014A&A...564A.125B}. This choice was motivated by the need to directly compute the Bayesian Evidence ($\mathcal{Z}$) of the models, which provides a more rigorous statistical metric for model comparison than the BIC approximation, particularly when combining heterogeneous datasets (RV and photometry) and exploring high-dimensional parameter spaces.

Based on our previous findings, we adopted a one-planet model for the combined modelling. This approach enabled us to derive consistent and precise estimates of the orbital and physical parameters of TOI-4602\,b by combining the complementary constraints from both transit and RV data. In particular, the joint fit yields a self-consistent determination of the planetary density, based on independently derived - yet coherently modelled - measurements of the planet’s radius and mass. Although the RV-only analysis strongly favours a circular orbit (M1), we decided to test the eccentric model also. This allows us to verify if the inclusion of the transit data introduces any preference for a non-circular orbit, and to derive more robust uncertainties for the planetary parameters by marginalising over the eccentricity. The planetary system parameters derived from the joint fit for both the circular and eccentric models are listed in Table \ref{tab:join_fit_results}.

We compared the models using the Bayes factor $\mathcal{B}_{ij} = \mathcal{Z}_i / \mathcal{Z}_j$. Following the scale of \citet{jeffreys1998theory}, a difference in log-evidence $\Delta \ln \mathcal{Z} > 2.5$ corresponds to moderate evidence, while $\Delta \ln \mathcal{Z} > 5$ indicates strong evidence in favour of the model with higher probability.
The eccentric model (using a uniform prior on $e$) yields a log-evidence of $\ln \mathcal{Z}_e = 468337.8 \pm 0.3$, compared to $\ln \mathcal{Z}_c = 468334.3 \pm 0.3$ for the circular case. This results in a Bayes factor of $\Delta \ln \mathcal{Z} = 3.5$ in favour of the eccentric solution ($e = 0.23 \pm 0.06$), corresponding to moderate evidence according to the scale of \citet{jeffreys1998theory}.  To ensure this preference was not an artifact of the prior, we performed additional fits using physically motivated distributions: a Beta distribution ($\alpha = 0.867, \beta = 3.03$; \citealt{Kipping2013}) and a Half-Gaussian prior ($\sigma = 0.32$), the latter being specifically recommended for small planets \citep{VanEylen2019AJ....157...61V}. These tests confirm the prior-independence of the result: the Beta prior yields $e = 0.217^{+0.055}_{-0.058}$ and $\Delta \ln \mathcal{Z} \approx 3.7$, while the Half-Gaussian prior results in $e = 0.224^{+0.056}_{-0.058}$ and $\Delta \ln \mathcal{Z} \approx 4.3$. Although the eccentric model is statistically favoured across all scenarios, the evidence remains below the "decisive" threshold ($\Delta \ln \mathcal{Z} < 5$). Furthermore, the planetary parameters remain consistent within $1\sigma$ regardless of the model choice; for instance, the eccentric fit yields a mass of $6.2 \pm 0.9 \, M_{\oplus}$, fully compatible with the circular case. 
Given the moderate significance of the improvement, the strong preference for circularity in the RV-only analysis, and the robustness of the planetary parameters against the model choice, we adopt the circular solution as our fiducial model. This provides a conservative description of the system with fewer free parameters, despite we cannot rule out a moderate eccentricity.

\begin{figure*}[h!]
	\centering
	\includegraphics[scale=0.42]{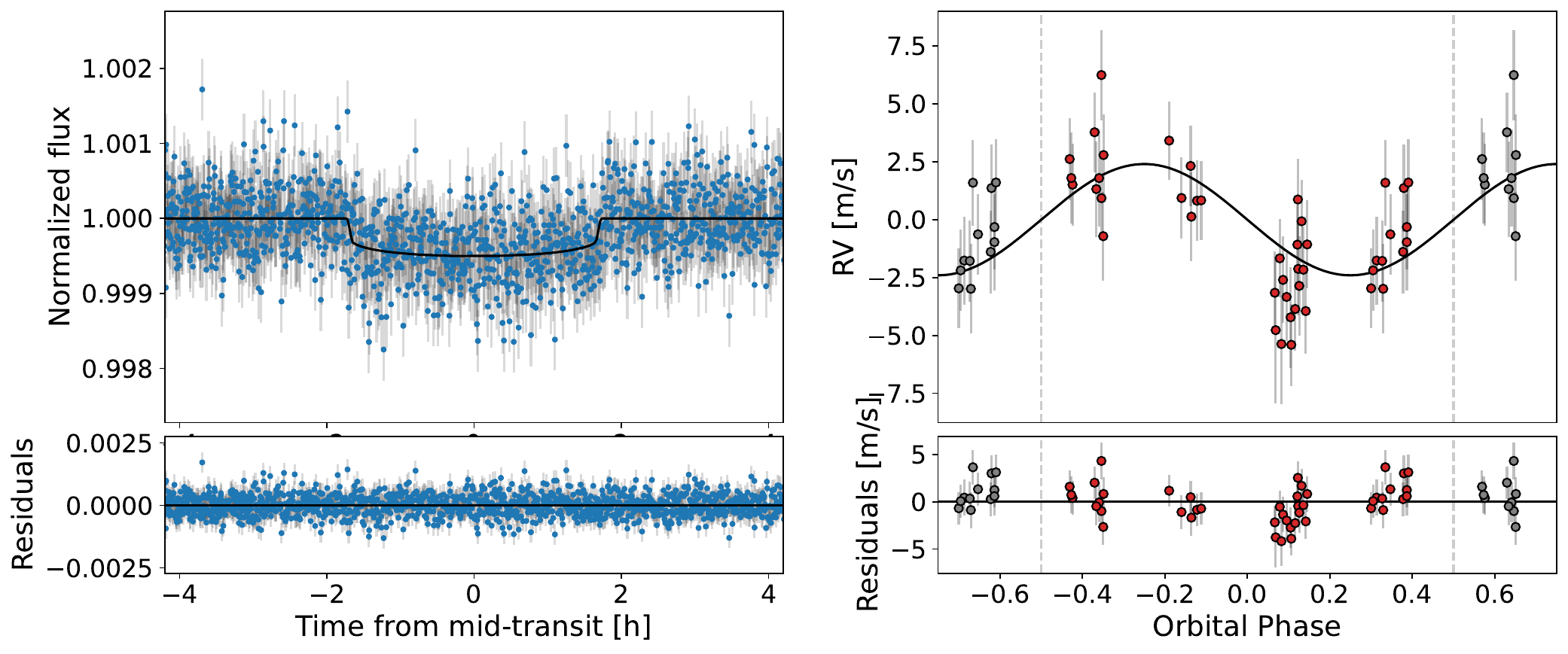}
	\caption{Modelling of photometric and spectroscopic time-series obtained from the joint fit assuming a circular model. (Left) Photometric modelling of the TOI-4602\,b planetary signal. The top panel displays the normalized, phase-folded TESS transits overlaid with the best-fit model (black line), while the bottom panel shows the residuals between the observed data and the model. (Right) Phase-folded RV fit of TOI-4602\,b planetary signal. The reported error bars include the jitter term, added in quadrature. The bottom panel displays the residuals of the fit.  }  
		\label{fig:joint_fit}
\end{figure*}

Fig. \ref{fig:joint_fit} shows the RV and transit data with the best-fit circular models overplotted. Based on this adopted solution, we determine that the transiting object is a sub-Neptune planet with a mass of $M_p = 5.5 \pm 0.9\,M_{\oplus}$ and a radius of $2.5 \pm 0.2 \,R_{\oplus}$, yielding a density of $\rho_p = 2.2 \pm 0.8$ g cm$^{-3}$. The planetary mass derived from the joint analysis is consistent with that obtained from our RV-only analysis. As expected, the combined fit allows a more precise determination of the orbital period of TOI-4602\,b ($P = 3.9812999^{+0.0000047}_{-0.0000048}$ days), in excellent agreement with the photometric period.

\subsection{Insolation, equilibrium temperature, scale height, and transmission spectroscopy metric} \label{sec:metric}
In addition to all the derived planet parameters, reported in Table \ref{tab:join_fit_results}, including the inclination, impact parameter, and semi-major axis, we calculated an insolation of 664 $\pm$ 42 $\rm S_\oplus$. 
We estimated the planet's equilibrium temperature (T$_{\mathrm{eq}}$) assuming a Bond albedo of A = 0.3 (the same as Earth's) and full heat redistribution between the day and night sides. This yields an equilibrium temperature of $T_{\mathrm{eq}} = 1344 \pm 80$ K.  
Assuming a cloud-free, H$_2$-dominated atmosphere, we estimated the atmospheric scale height to be $H = 538\pm140$ km.
This corresponds to a transmission spectral feature amplitude of 60$\pm$20 ppm, estimated using the relation $\delta_\lambda \sim 4R_pH/R_\star^2$ \citep{Kreidberg2018haex.bookE.100K}.  
We note that the assumption of a hydrogen-dominated atmosphere represents the most optimistic case for transmission spectroscopy. For super-Earths, heavier atmospheric compositions (e.g dominated by water vapour or CO$_2$) are also plausible and would imply higher mean molecular weights. In such scenarios, the scale height-and consequently the amplitude of spectral features-would be significantly reduced, making atmospheric characterization more challenging.
We also computed the Transmission Spectroscopy Metric (TSM) following the definition by \citet{Kempton2018PASP..130k4401K}. We obtained a value of TSM = 140 $\pm$ 54, which lies above the threshold of 90 suggested by \citet{Kempton2018PASP..130k4401K}, indicating that the planet is a promising target for atmospheric characterization.

\section{Mass-radius diagram and internal structure}\label{sec:internal}
In this section we first use our measurements of the planet parameters to investigate the bulk structure of TOI-4602\,b in the context of the photo-evaporation as the main driver of the evolution of these close-in, low-mass planets.

Figure \ref{fig:massa-radius-diagram} shows the position of TOI-4602\,b in a mass-radius diagram, along with a sample of well-characterized exoplanets with a mass uncertainty better than 20\% and a radius uncertainty better than 10\%. The parameters for the known exoplanets were taken from NASA Exoplanet Archive \citep{Christiansen2025PSJ.....6..186C}. We included theoretical mass-radius relationships for different bulk compositions: bare rocky planets with 100\% Fe, 33\% Fe + 66\% MgSiO$_3$, 20\% Fe + 80\% MgSiO$_3$, 100\% MgSiO$_3$ and Earth-like rocky core with H/He envelopes of 0.3\% and 1\% by mass. These models assume  a surface pressure of 1 mbar and an equilibrium temperature of 1000 K. Although models with high amounts of surface water of tens of weight percents could in principle fit TOI-4602\,b, we refrain from plotting such models as they are not plausible given interior-atmosphere interactions \citep{Luo2024NatAs...8.1399L,Werlen2025ApJ...991L..16W}. Specifically, global equilibrium calculations predict a universal upper limit on the available surface water of few percent \citep{Werlen2025ApJ...991L..16W}, even if large amounts of water are accreted during formation. The majority of accreted water is chemically destroyed and stored in the metal phase of sub-Neptunes \citep{schlichting2022chemical}. Sub-Neptunes are thus hosting enriched H/He envelopes, chemically coupled to their deeper magma oceans.

In order to determine the possible interior compositions and structures, we employ an inference framework \citep{DeWringer} using the physical forward model of \cite{Dorn2015A&A...577A..83D,Dorn2017A&A...597A..37D} with recent updates from \cite{dorn_hidden_2021, Luo2024NatAs...8.1399L}. The inference framework is a surrogate-accelerated Bayesian inference, in which the computationally expensive physical forward model is replaced with a fast polynomial chaos-Kriging (PCK) surrogate directly within an MCMC sampling loop \citep{marelli2014uqlab}. For this inference, both the circular and eccentric models were tested. The surrogate model provides high quality fits with R-squared values (coefficient of determination). For the circular model, the fits are 0.9999 and 0.9849 for the planetary mass and radius, respectively, and for the eccentric model, the fits are 0.9999 and 0.9862 for the planetary mass and radius, respectively. Also, the surrogate root mean square errors are well below observational uncertainties with the same values for both models: 0.0006 and 0.06 for planetary mass and radius, respectively. Those errors of the model uncertainty are accounted for in the likelihood function.

The interior model assumes an Earth-like rocky interior with an iron-dominated core, a silicate mantle and an enriched H/He envelope on top. The core-mantle mass ratio is set to 0.325:0.675.
We allow for both liquid and solid phases in mantle and core layers.
For liquid core iron and iron alloys we use the equation of state (EOS) by \citet{Luo2024NatAs...8.1399L}. For solid iron, we use the EOS for hexagonal close packed iron \citep{hakim_new_2018,miozzi_new_2020}. 
For pressures below $\approx 125\,$GPa, the solid mantle mineralogy is modelled using the thermodynamical model \textsc{Perple\_X} \citep{connolly_geodynamic_2009} considering the system of MgO, SiO$_{2}$, and FeO. At higher pressures we define the stable minerals \textit{a priori} and use their respective EOS from various sources \citep{hemley_constraints_1992,fischer_equation_2011,faik_equation_2018,musella_physical_2019}.
The liquid mantle is modelled as a mixture of Mg$_2$SiO$_4$, SiO$_2$ and FeO \citep{melosh_hydrocode_2007,faik_equation_2018,ichikawa_ab_2020,Stewart2020AIPC.2272h0003S}, and mixed using the additive volume law. We assume an adiabatic temperature profile for the core and mantle with possible temperature jumps at the core-mantle-boundary depending on melt temperatures.

The H$_2$-He-H$_2$O atmosphere layer is modelled using the analytic description of \citet{Guillot2010A&A...520A..27G} and consists of an irradiated layer on top of a non-irradiated layer in radiative-convective equilibrium. 
The water mass fraction is given by the metallicity $Z$ and the hydrogen-helium ratio is set to solar. 
The two components of the atmosphere, H$_2$/He and H$_2$O, are again mixed following the additive volume law and using the EOS by \citet{1995_Saumon_EOS} for H$_2$/He and the ANOES EOS \citep{1990_thompson_aneos} for H$_2$O.
The transit radius of a planet is defined where the chord optical depth reaches 0.56 \citep{des2008rayleigh}. For each model realization, the planet intrinsic luminosity is calculated following \citep{mordasini_planetary_2020} and is a function of planet mass, atmospheric mass fraction and age.

While fixing an Earth-like core-mantle ratio, we find that the planet is best described by an envelope mass of $0.05 M_\mathrm{\oplus} \pm 0.03$ for both models. The metallicity is found to be $0.61^{+0.24}_{_0.26}$ for the circular model, and $0.60^{+0.24}_{_0.27}$ for the eccentric one. These results are visible in Fig. \ref{fig:inversion} and Table \ref{tab:priors_posteriors_combined}. Although we obtain two different planetary masses, they are compatible at 1-sigma and lead to the same scientific results. This scenario is compatible with the expected atmospheric loss as described in the next section.

\section{Photoevaporation and long-term evolution}\label{sec:fotoevaporazione}

We investigated TOI-4602\,b’s long-term atmospheric evolution, considering the significant UV-driven mass loss typical of close-in sub-Neptunes, and estimated how much atmosphere the planet may have retained over millions of years.

\begin{figure}
\centering
   \includegraphics[width=0.9\linewidth]{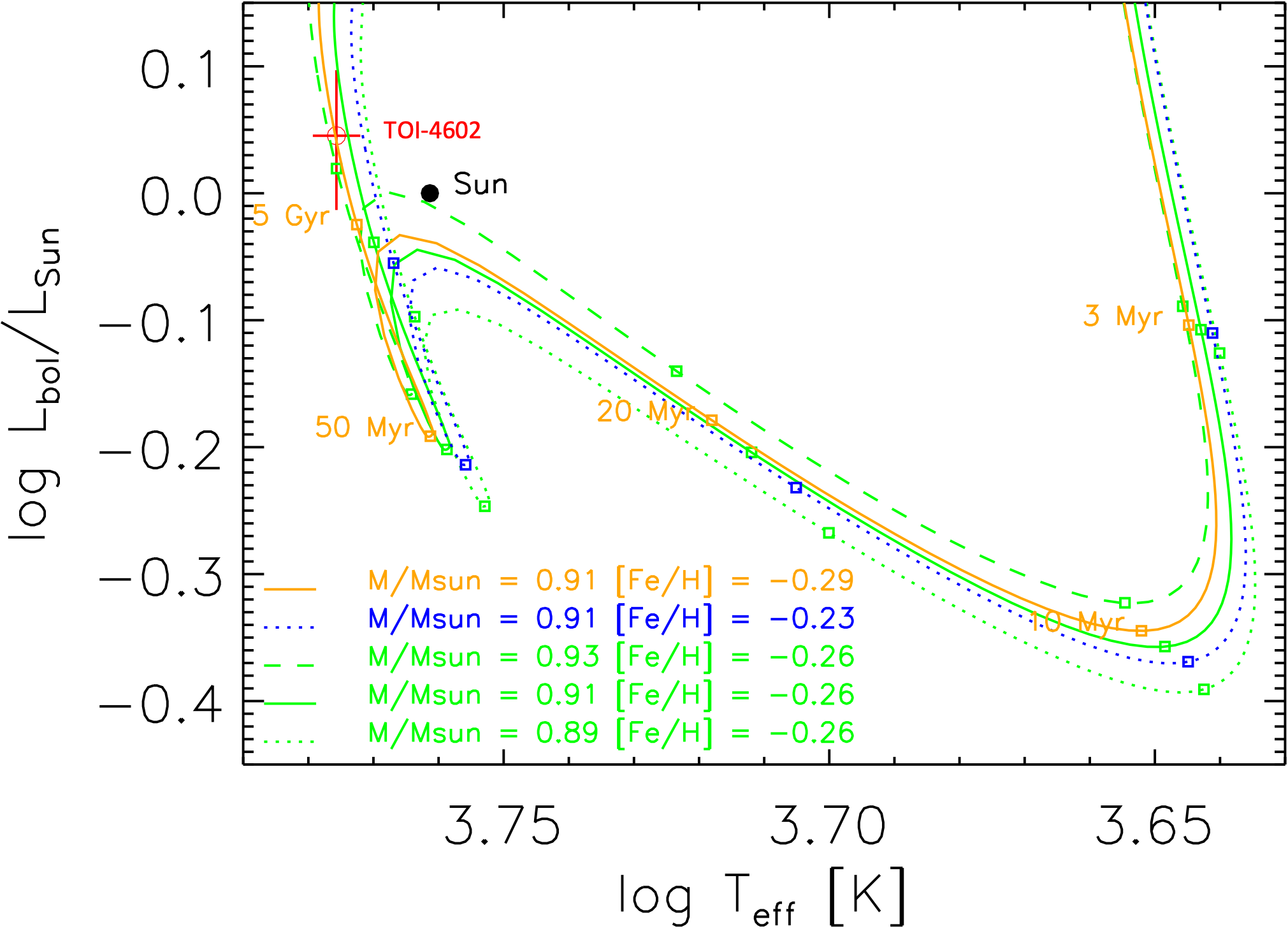}
\caption{Evolutionary track of TOI-4602 in the effective temperature-bolometric luminosity plane. The red circle marks the current location of the star on the track.} 
\label{fig:track_MESA}
\end{figure}

\begin{figure}
\centering
\includegraphics[width=0.9\linewidth]{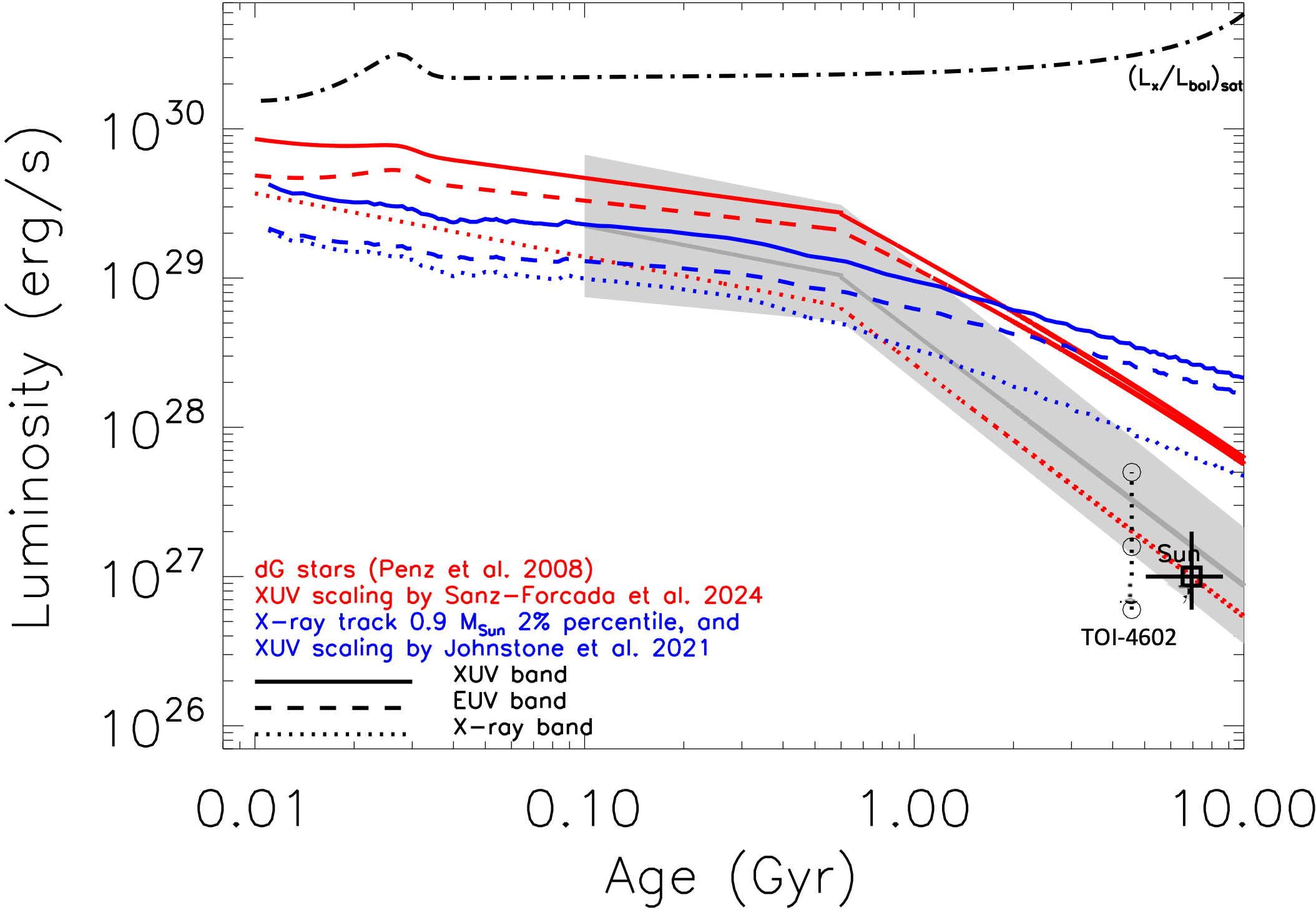}
\caption{Time evolution of X-ray ($5-100$\,\AA), EUV ($100-920$\,\AA), and total XUV luminosity of TOI-4602, according to \citet{Penz+2008} and the X-ray/EUV scaling by \citet{Sanz-Forcada+2025} (red lines) and according to \citet{Johnstone+2021} (blue lines). Uncertainties on the age and X-ray luminosity of TOI-4602 (square symbol) are also indicated, with respect to the nominal value of $L_{\rm x} = 10^{27}$\,erg/s at 6.8\,Gyr. The gray area is the original locus for dG stars in \citet{Penz+2008}. For comparison purpose, we also show the current position of the Sun and the full range of variability during an 11-year magnetic cycle.} 
\label{fig:LXUVevol}
\end{figure}

We selected a MESA evolutionary track \citep{Choi+2016}, compatible with the star's position at its nominal age in the theoretical L$_{\rm bol}$ vs.\ T$_{\rm eff}$ diagram. 
We considered a grid of evolutionary tracks for stars with masses and metallicities equal to the nominal values for TOI-4602, or equal to $\pm 1\sigma$ values in both parameters (9 tracks in total).
Some of them are shown in Fig.~\ref{fig:track_MESA}.
We found that the best match is achieved with the track of a star with the nominal mass of TOI-4602 and metallicity [Fe/H]$ = -0.29$ ($-1\sigma$).
We used this track for modelling the bolometric irradiation of the planet and therefore its equilibrium temperature, as well as the core-envelope structure during evolution, taking into account the gravitational shrinking \citep{Fortney+2007}.
To model the evolution of high-energy irradiation, we assumed as anchor point for the X-ray luminosity at the current age the value $L_{\rm x} = 1 \times 10^{27}$\,erg/s, determined in Sect.\ \ref{sec:activity}.

In Fig.~\ref{fig:LXUVevol}  we show the time evolution of the X-ray ($5-100$\,\AA), EUV ($100-920$\,\AA), and XUV (X+EUV) luminosities, according to two different models: the \citet{Penz+2008} X-ray luminosity evolution coupled with the \citet{Sanz-Forcada+2025} X-ray to the EUV scaling law (PSF model) or, alternatively, adopting the \citet{Johnstone+2021} prescription (Jo model). 
In the former case, the current X-ray luminosity appears to be near the median for solar-mass stars at the nominal age of TOI-4602.
In the second case, we selected the evolutionary track corresponding to the lowest $2\%$ percentile of the activity distribution appropriate for stars with a mass of $0.9\,M_{\sun}$. 
Nonetheless, the \citet{Johnstone+2021} evolutionary path yields an X-ray luminosity that is about a factor 6 higher than the nominal value determined above, and significantly outside the uncertainty range. Hence, with the Jo-model assumption, the high-energy irradiation of the planet for ages $> 800$\,Myr is likely overestimated.

Considering the constraints on the activity level provided by the age and the chromospheric Ca II H\&K index (Sect.\ \ref{sec:activity}, we also explored the results of photo-evaporation assuming the \citet{Penz+2008} evolutionary tracks anchored at the extreme values of the uncertainty range : $L_{\rm x} = 2 \times 10^{27}$\,erg/s at an age of 5\,Gyr, or $L_{\rm x} = 6 \times 10^{26}$\,erg/s at an age of 8.6\,Gyr). 

In summary, we investigated the planetary photoevaporation with four evolutionary tracks for the high-energy (X-ray + EUV) radiation: two tracks in which we fixed the stellar age at 6.8\,Gyr, and we assumed either the PSF model or the Jo model (PSF-6.8 or Jo-6.8, respectively), and two more tracks at ages of 5\,Gyr or 8.6\,Gyr, both following the PSF model (PSF-5 and PSF-8).
In Fig. \ref{fig:evap} we show the evolution of the mass, radius, and mass loss rate for these four different scenarios.

For the analysis of the past and future evaporative evolution, we employed the numerical code first presented in \citet{Locci19}, subsequently adapted for the study of single systems \citep{Maggio22} and more recently described in \cite{Mantovan24b}. 
Assuming an Earth-like core composition, we first calculated the planet’s internal structure, finding a core mass of approximately $5.5\,M_\oplus$ and a radius of around $1.56\,R_\oplus$, which implies an atmospheric fraction of approximately 1.5\%. These values are quite insensitive to variations in age. This finding is in good agreement with the results of section \ref{sec:internal}.
We then computed the past and future evolution of key planetary parameters, such as the atmospheric fraction and total radius, while keeping the mass and core radius constant over time. In order to calculate the mass-loss rates we use the ATES analytical approximation, which includes different regimes such as energy limited and recombination limited
 \citep{Caldiroli21}.
Assuming solar abundances for the atmosphere, we followed the atmospheric evolution back to an age of 10 million years, which we assumed to be the time when the accretion disc was fully dissipated and the planet had reached its final orbit, and forward in time up to an age of 10 Gyr. 
The calculated initial mass and radius of the system depend on the XUV evolutionary track used. 
Overall, for the initial mass we predict values in the range 31.1 -- 33.8\,$M_\oplus$, with corresponding radii 14.8 -- 14.7\,$R_\oplus$.

Furthermore, we found that for the PSF-6.8. Jo-6.8 and PSF-5 tracks, the planet is currently undergoing the transition across the radius valley. In these scenarios, it will lose its entire remaining atmosphere within approximately 2.6, 0.8, and 1.4 Gyr from now, respectively (i.e., reaching a bare-core state at an age of about 9.4, 7.6, and 6.4 Gyr). Therefore, under these assumptions, the planet will eventually cross the radius valley entirely, ending its evolution on the small-radius peak of the bimodal distribution with a mass and radius equal to that of its core.
In contrast, for the PSF-8 track (which assumes a current age of 8.6 Gyr and a milder XUV history), the planet is currently at the end of its active evaporative phase. It is able to retain a small atmospheric fraction against further photoevaporation, meaning its radius will stall near its current value of about $2 R_\oplus$. In this scenario, the planet's evolution halts just above the radius valley, remaining part of the sub-Neptune population.

We also explored a case where we assumed a metallicity 50 times that of the Sun for the calculation of the envelope radius \citep{LopFor14}, finding similar behaviour to the previous case for all four tracks, in particular in the future evolution. The differences between the two scenarios are more significant in the initial phases, where the planet exhibits a larger radius, in the case of high metallicity. Consequently, it is subject to greater evaporation, which in turn requires a higher initial mass that ranges from 38 to 42 $M_\oplus$, depending on the evolutionary track considered. The time evolution of planetary parameters for the enhanced abundance case is shown in Fig. \ref{fig:evap50x}. 

Finally, we simulated the case with eccentric orbit, resulting in a higher planetary density compared to the circular-orbit scenario. As a consequence, TOI-4602\,b has a slightly greater resistance to photoevaporation. Although the effect is modest, it implies that in the three cases in which the planet completely lost its atmosphere in the circular-orbit scenario, the atmosphere is still lost but at slightly later ages. In particular, this occurs at about 10 Gyr for the PSF-6.8 case, 7.8 Gyr for the Jo-6.8 case, and 6.7 Gyr for the PSF-5 case. Otherwise, the evolutionary curves remain similar to those of the circular-orbit case.

In conclusion, we find that the planet's final radius is largely independent of the assumed initial atmospheric metallicity, although higher metallicities strongly impact the early evolutionary phases. Crucially, the current evolutionary stage and the ultimate fate of TOI-4602 b are highly sensitive to the stellar age and the assumed XUV history. In three out of four analysed XUV scenarios, the planet is currently still undergoing atmospheric escape and will eventually cross the radius valley, ending as a bare core on the small-radius peak. The only exception is the oldest scenario (8.6 Gyr) coupled with the PSF-8 XUV track; here, the planet has already ceased significant evaporation and its evolution will safely terminate above the radius valley ($\sim 2 R_\oplus$), retaining a thin envelope.

\section{Atmospheric simulations}\label{sec:ariel_retrieval}
TOI-4602\,b is an excellent candidate for atmospheric characterization due to its high TSM (Section \ref{sec:metric}) and the brightness and quietness of its host star. Here, we investigate the capability of Ariel and JWST \citep{Gardner2006SSRv..123..485G} to characterize its atmosphere via retrievals on simulated 1$\times$ solar-metallicity transmission spectra. We assumed a cloud-free, isothermal, plane-parallel atmosphere with constant molecular abundances, including Rayleigh scattering and Collision Induced Absorption (CIA) of H$_2$-H$_2$/H$_2$-He \citep{Richard2012JQSRT.113.1276R, cox2015allen}. Forward and retrieval models utilize ExoMol \citep{Tennyson2016JMoSp.327...73T} and HITEMP \citep{Rothman2008hitr.confE..14R} opacities for H$_2$O \citep{Polyansky2018MNRAS.480.2597P}, CO \citep{Li2015ApJS..216...15L}, CO$_2$ \citep{Yurchenko2020MNRAS.496.5282Y}, CH$_4$ \citep{Yurchenko2024MNRAS.528.3719Y}, SO$_2$ \citep{Underwood2016MNRAS.459.3890U}, and K \citep{Allard2016A&A...589A..21A}. Although CH$_4$ likely dissociates into CO above $\sim$1000 K, we included it to probe potential disequilibrium chemistry and constrain upper limits.

Ariel simulations were performed using the TauREx framework \citep{Al-Refaie2021ApJ...917...37A, Al-Refaie2022ApJ...932..123A}. Forward spectra, based on parameters in Tables \ref{tab:stellar_parameters} and \ref{tab:join_fit_results}, were convolved with the ArielRad instrument model (v. 2.4.26, \citealt{Mugnai2020ExA....50..303M}; Payload v. 0.0.17, ExoRad v. 2.1.111, \citealt{Mugnai2023JOSS....8.5348M}) to generate input spectra. We focused on Tier 2 resolution \citep{Tinetti2022EPSC...16.1114T}, binning the data to R = 10, 50, and 20 in FGS-NIRSpec (1.1-1.95 µm), AIRS-Ch0 (1.95-3.9 µm), and AIRS-Ch1 (3.9-7.8 µm), respectively\footnote{The Ariel payload will consist of two instruments: the FGS providing the Fine Guidance System capabilities combined with a VIS photometer and a NIR spectrometer, and the Ariel IR Spectrometer (AIRS) providing spectra in the $1.95\text{--}7.8\,\mu\text{m}$ range with a resolution of 30–100.}. Using ArielRad, we determined that 20 transits are required to achieve the Tier-2 S/N, assuming Gaussian noise scaled by the square root of observations. Retrievals employed the \textsc{PyMultinest} sampler with 500 live points, a 0.5 evidence tolerance, and non-informative uniform priors on abundances.

We simulated and retrieved JWST observations using the \textsc{Pyrat Bay} framework \citep{CubillosBlecic2021mnrasPyratBay}, adopting the same atmospheric assumptions as the Ariel case. Using the \textsc{Gen~TSO} simulator \citep{Cubillos2024paspGenTSO}, we generated transit depths and uncertainties for a single transit combining NIRISS/SOSS (SUB204STRIPE), NIRSpec/BOTS (G395H), and MIRI/LRS to cover 0.65-12\,{\micron}. Retrievals employed the \textsc{PyMultinest} sampler. The free parameters included isothermal temperature, reference pressure at $R_{\rm p}$, gray cloud deck pressure, and VMRs for K, \ch{H2O}, \ch{CO2}, CO, \ch{CH4}, and \ch{SO2}.

Fig. \ref{fig:atmo_simulations} shows the simulated transmission spectra and resulting posterior distributions for the Ariel (blue) and JWST (orange) retrieval runs. While both recover the main atmospheric constituents, differences in spectral coverage and resolution drive key discrepancies in precision.
Ariel achieves higher precision on H$_2$O ($\sim$0.2 dex vs $\sim$0.5 dex) and SO$_2$ due to its broad, continuous wavelength coverage (0.5–7.8 $\mu$m), which captures multiple water bands and the strong SO$_2$ feature at 7.3 $\mu$m, effectively breaking parameter degeneracies. Conversely, JWST provides superior constraints on Potassium (K). In the Ariel simulations, the K-doublet signal ($\sim$ 0.77 $\mu\text{m}$) falls in the blue end of the spectrum, where it is constrained by only a few photometric channels. Due to this limited spectral information, the narrow K absorption feature is highly degenerate with the spectral slope introduced by Rayleigh scattering, cloud opacity, and stellar activity. Since these strong contaminating effects are likely to be present in real observational data, the constraint on K is significantly compromised. Consequently, the resulting broad posterior is considered difficult to confidently interpret as a true atmospheric detection, highlighting the value of complementary observations between Ariel and JWST.
Regarding carbon species, CO accuracy is comparable, and CH$_4$ is correctly retrieved as an upper limit in both cases. For CO$_2$ we noticed that the true value is slightly outside the 1$\sigma$ region for both retrievals. This likely results from the relatively low CO$_2$ abundance at 1$\times$ solar metallicity, and the blending of multiple opacities (CO, CO$_2$, H$_2$O, SO$_2$) at around  4.0 $\mu$m, the region with the strongest CO$_2$ absorption band. Despite the challenges linked to degeneracies in the Potassium band, the Ariel high-precision constraints on broadly distributed key species like H$_2$O and SO$_2$ demonstrates the robustness and effectiveness of its wide-field survey approach for characterising a large sample of exoplanet atmospheres.

   
%


%

%

\section{Discussion and Conclusions}
\label{sec:discussion_conclusions}

We presented a comprehensive analysis of the TOI-4602 system, combining TESS photometry with high-precision radial velocities from HARPS-N obtained within the ArMS project. Our joint analysis tested both circular and eccentric orbital models, confirming the presence of a transiting planet with a period of 3.98 days. While both models yield consistent planetary mass and radius estimates, the Bayesian evidence shows a moderate preference for the eccentric solution. However, interpreting this eccentricity requires a consistent physical model that accounts for the system's advanced age (8-9 Gyr) and the planet's internal composition.
The circularization timescale of the orbit of TOI-4602\,b is highly uncertain because we do not know its modified tidal quality factor $Q^{\prime}$ that quantifies the efficiency of tidal dissipation in its interior.
If we adopt a value of $Q^{\prime} = 10^{5}$, that is similar to that of Uranus \citep[see][]{Ogilvie2014ARA&A..52..171O}, and use the tidal model of \citet{Leconte2010A&A...516A..64L} computing the product of the tidal time lag $\Delta t$ by the Love number $k_{2}$ with the simple formula $k_{2} \Delta t = 3/(2Q^{\prime} n)$, where $n = 2\pi/P_{\rm orb}$ is the orbital mean motion, we obtain a tidal decay timescale $\tau_{\rm e} \equiv |e/\dot{e}| \sim 3.4$ Gyr. This timescale is shorter, but still comparable with the age of the system. Therefore, in this scenario, a remarkable primordial eccentricity could not have been completely damped during the main-sequence lifetime of the system. On the other hand, if we assume that the planet has an inner rocky core that dominates tidal dissipation, surrounded by a thick hydrogen envelope that is not appreciably contributing to tidal dissipation, the situation changes. Considering a rocky core of radius $1.56$~R$_{\oplus}$ (cf. Sect. \ref{sec:fotoevaporazione}) and $Q^{\prime} = 10^{3}$, we obtain, we obtain a tidal damping timescale of 0.2 Gyr, much shorter than the age of the star. In this scenario, any primordial eccentricity has certainly been damped at the present evolutionary stage of the system. These estimates are valid assuming no additional body in the system able to gravitationally excite the eccentricity of TOI-4602\,b.
Our internal structure modelling revealed that TOI-4602\,b is indeed best described by a rocky core surrounded by a tenuous envelope ($\sim1\%$ by mass). This composition strongly points towards the low-$Q^{\prime}$ scenario ($\tau_{\rm e} \sim 0.2$ Gyr), rendering the survival of primordial eccentricity physically implausible.

To rigorously assess the presence of perturbers, we performed an injection-recovery test combining HARPS-N RVs and proper motion anomalies, finding sensitivity to Jupiter-mass planets up to $\sim 1$ AU. These observational constraints effectively rule out secular perturbations as the cause of the eccentricity, since generating $e \approx 0.23$ would require a massive companion at $\sim 0.11$ AU or a compact chain at $0.07-0.08$ AU, which would have been detected with high confidence. Alternative scenarios involving distant companions, such as high-eccentricity migration from $> 1$ AU or planet-planet scattering, remain unconstrained by our data but are physically disfavoured. The former requires a high tidal quality factor inconsistent with the planet's likely rock-dominated structure, while the latter would require a statistically unlikely recent instability given the fast tidal damping ($\tau_e \sim 0.2$ Gyr) expected for a rocky core.

Given the conflict between the rapid tidal damping expected for a rocky planet and the lack of detected close-in perturbers, we adopt the circular solution as our fiducial model, classifying TOI-4602\,b as a sub-Neptune with a mass of $M_p = 5.5 \pm 0.9 \, M_{\oplus}$ and a radius of $R_p = 2.5 \pm 0.2 \, R_{\oplus}$, yielding a mean bulk density of $\rho_p = 2.2 \pm 0.8$ g cm$^{-3}$. 

TOI-4602\,b currently resides in the sub-Neptune regime, populating the upper peak of the bimodal radius distribution. With a radius of $2.5 \pm 0.2 \, R_{\oplus}$ and a stellar insolation of $664 \pm 42 \, S_{\oplus}$, the planet sits well above the radius valley \citep{2017AJ....154..109F}. This position is particularly interesting when considering the recent findings of \citet{Wanderley2025ApJ...993..233W}, who revisited the dependence of the valley on orbital period and insolation. At such high irradiation levels and short orbital periods ($P \sim 3.98$ d), many planets are expected to cross the valley due to atmospheric stripping, becoming bare rocky cores. 

Our photo-evaporation modelling (Section \ref{sec:fotoevaporazione}) suggests that TOI-4602\,b is a "survivor" in this high-insolation regime. Despite the intense hydrodynamic escape expected during the system's youth, the planet has retained a non-negligible H/He envelope. Depending on the stellar XUV history, TOI-4602\,b is either in a slow phase of atmospheric mass loss or has reached a stable configuration, offering a valuable snapshot of a sub-Neptune that has resisted complete transformation into a bare rocky core. This resilience provides a critical constraint for models of atmospheric retention and evolution for planets transitioning across the radius valley.

Finally, with a Transmission Spectroscopy Metric (TSM) of $140 \pm 54$, TOI-4602\,b is a prime target for atmospheric characterization. Our retrieval simulations demonstrate that Ariel and JWST can effectively constrain its atmospheric composition.
Spectroscopic measurements are crucial to probe the chemical inventory and thermal structure of a planet transitioning across the radius valley. Critically, resolving the abundance of species like H$_2$O and placing limits on the envelope metallicity would provide an independent check on our internal structure assumptions. Confirming a low-mass, high-metallicity envelope would validate the low-$Q'$ hypothesis, definitively closing the loop on the dynamical history of this evolved planetary system. TOI-4602\,b thus stands out as a key laboratory where atmospheric characterization can directly illuminate the mechanisms sculpting the architecture of the small-planet population.

\begin{acknowledgements}
The authors acknowledge the support from ASI-INAF agreement 2021-5-HH.0, the INAF Guest Observer Grant (Normal) "ArMS: the Ariel Masses Survey Large Program at the TNG" (Prog ID: \texttt{AOT48TAC\_48}), and the INAF Mini-Grant "Impact of planetary Masses and Radii Estimates on the Atmospheric retrievals - IMaREA". The research activities described in this paper were carried out with contribution of the Next Generation EU funds within the National Recovery and Resilience Plan (PNRR), Mission 4 - Education and Research, Component 2 - From Research to Business (M4C2), Investment Line 3.1 - Strengthening and creation of Research Infrastructures, Project IR0000034 – “STILES - Strengthening the Italian Leadership in ELT and SKA”. C.D. acknowledges support from the Swiss National Science Foundation under grant TMSGI2\_211313. This work has been carried out within the framework of the NCCR PlanetS supported by the Swiss National Science Foundation under grant 51NF40\_205606. G.M. acknowledges support by the Space It Up project funded by the Italian Space Agency, ASI, and the Ministry of University and Research, MUR, under contract n. 2024-5-E.0 - CUP n. I53D24000060005. We acknowledge the Italian center for Astronomical Archives (IA2), part of the Italian National Institute for Astrophysics (INAF), for providing technical assistance, services and supporting activities of the GAPS collaboration. This work used data from the European Space Agency (ESA) mission {\it Gaia} (\url{https://www.cosmos.esa.int/gaia}), processed by the {\it Gaia} Data Processing and Analysis Consortium (DPAC,
\url{https://www.cosmos.esa.int/web/gaia/dpac/consortium}). Funding for DPAC has been provided by national institutions participating in the {\it Gaia} Multilateral Agreement. 
\end{acknowledgements}

\balance{
    \bibliographystyle{aa}
    \bibliography{bibliography}}


\appendix

\onecolumn

\section{Data: additional figures and tables} 

\begin{table}[ht!]													
\footnotesize															
\centering																
\caption{HARPS-N time series of TOI-4602.}				
\label{tab:timeseries}											
\begin{tabular}{lcccc}												
\hline																
\hline																
\rule{0pt}{2.8ex}BJD - 2450000 & RV & BIS-SPAN   & $\rm \log R'_{HK}$ & H$\alpha$ \\		
& (m s$^{-1}$) & (m s$^{-1}$) & (m s$^{-1}$) &  \\
\hline																	
\rule{0pt}{2.8ex}10592.70	&	-48511.6	$\pm$	0.7	&	5.6	&	-4.964	$\pm$	0.003	&	0.1779	$\pm$	0.0002	\\	
10594.73	&	-48505.2	$\pm$	0.6	&	5.7	&	-4.965	$\pm$	0.003	&	0.1778	$\pm$	0.0002	\\	
10596.73	&	-48510.4	$\pm$	1.1	&	3	&	-4.963	$\pm$	0.007	&	0.1784	$\pm$	0.0003	\\	
10600.77	&	-48510.4	$\pm$	0.8	&	5.5	&	-4.962	$\pm$	0.004	&	0.1796	$\pm$	0.0002	\\	
10607.77	&	-48511.2	$\pm$	1.0	&	4	&	-4.976	$\pm$	0.007	&	0.1791	$\pm$	0.0002	\\	
10612.74	&	-48507.0	$\pm$	1.0	&	9	&	-4.975	$\pm$	0.006	&	0.1789	$\pm$	0.0002	\\	
10613.70	&	-48507.1	$\pm$	0.8	&	9	&	-4.984	$\pm$	0.006	&	0.1793	$\pm$	0.0003	\\	
10617.65	&	-48506.0	$\pm$	0.8	&	7	&	-4.971	$\pm$	0.005	&	0.1789	$\pm$	0.0002	\\	
10618.73	&	-48507.7	$\pm$	0.8	&	5	&	-4.976	$\pm$	0.005	&	0.1783	$\pm$	0.0002	\\	
10619.68	&	-48510.3	$\pm$	0.6	&	9.9	&	-4.970	$\pm$	0.003	&	0.1786	$\pm$	0.0002	\\	
10620.68	&	-48511.6	$\pm$	1.1	&	6	&	-4.985	$\pm$	0.008	&	0.1788	$\pm$	0.0003	\\	
10625.63	&	-48506.8	$\pm$	1.2	&	8	&	-4.974	$\pm$	0.009	&	0.1790	$\pm$	0.0004	\\	
10643.59	&	-48514	$\pm$	2	&	13	&	-4.97	$\pm$	0.02	&	0.1785	$\pm$	0.0005	\\	
10652.60	&	-48509.3	$\pm$	0.7	&	3.4	&	-4.967	$\pm$	0.004	&	0.1781	$\pm$	0.0002	\\	
10655.47	&	-48511.8	$\pm$	0.9	&	6	&	-4.956	$\pm$	0.005	&	0.1788	$\pm$	0.0003	\\	
10656.42	&	-48510.8	$\pm$	0.8	&	1.7	&	-4.972	$\pm$	0.005	&	0.1785	$\pm$	0.0002	\\	
10671.55	&	-48514.1	$\pm$	0.8	&	8	&	-4.948	$\pm$	0.005	&	0.1788	$\pm$	0.0002	\\	
10674.57	&	-48508.5	$\pm$	1.1	&	8	&	-4.950	$\pm$	0.008	&	0.1793	$\pm$	0.0003	\\	
10675.53	&	-48512.9	$\pm$	1.5	&	17	&	-4.968	$\pm$	0.013	&	0.1783	$\pm$	0.0004	\\	
10682.52	&	-48506.3	$\pm$	0.8	&	11.6	&	-4.940	$\pm$	0.004	&	0.1788	$\pm$	0.0002	\\	
10683.45	&	-48512.0	$\pm$	1.1	&	6	&	-4.954	$\pm$	0.007	&	0.1789	$\pm$	0.0003	\\	
10687.54	&	-48510.8	$\pm$	1.2	&	1	&	-4.951	$\pm$	0.009	&	0.1793	$\pm$	0.0003	\\	
10687.55	&	-48511.5	$\pm$	1.5	&	12	&	-4.942	$\pm$	0.012	&	0.1788	$\pm$	0.0004	\\	
10688.56	&	-48507.3	$\pm$	1.1	&	2	&	-4.949	$\pm$	0.007	&	0.1786	$\pm$	0.0003	\\	
10691.57	&	-48510.8	$\pm$	1.0	&	5	&	-4.935	$\pm$	0.007	&	0.1787	$\pm$	0.0003	\\	
10692.59	&	-48507.0	$\pm$	1.0	&	3	&	-4.944	$\pm$	0.007	&	0.1785	$\pm$	0.0003	\\	
10693.58	&	-48506.8	$\pm$	1.4	&	5	&	-4.956	$\pm$	0.012	&	0.1789	$\pm$	0.0003	\\	
10695.48	&	-48512.5	$\pm$	0.9	&	7	&	-4.952	$\pm$	0.005	&	0.1793	$\pm$	0.0003	\\	
10701.50	&	-48504.9	$\pm$	0.7	&	8.0	&	-4.944	$\pm$	0.004	&	0.1786	$\pm$	0.0002	\\	
10703.50	&	-48508.7	$\pm$	0.8	&	6	&	-4.938	$\pm$	0.005	&	0.1777	$\pm$	0.0002	\\	
10704.48	&	-48510.0	$\pm$	0.8	&	8	&	-4.947	$\pm$	0.005	&	0.1781	$\pm$	0.0002	\\	
10715.40	&	-48509.7	$\pm$	0.7	&	3.7	&	-4.954	$\pm$	0.004	&	0.1786	$\pm$	0.0002	\\	
10717.49	&	-48507.7	$\pm$	0.8	&	3	&	-4.921	$\pm$	0.004	&	0.1784	$\pm$	0.0002	\\	
10718.42	&	-48507.8	$\pm$	0.8	&	7.2	&	-4.941	$\pm$	0.004	&	0.1786	$\pm$	0.0002	\\	
10719.47	&	-48512.6	$\pm$	0.9	&	5	&	-4.941	$\pm$	0.006	&	0.1786	$\pm$	0.0003	\\	
10720.44	&	-48509.0	$\pm$	0.9	&	9	&	-4.938	$\pm$	0.006	&	0.1785	$\pm$	0.0003	\\	
10721.48	&	-48502.4	$\pm$	1.1	&	3	&	-4.926	$\pm$	0.008	&	0.1796	$\pm$	0.0003	\\	
10723.37	&	-48507.8	$\pm$	0.8	&	9.9	&	-4.944	$\pm$	0.004	&	0.1788	$\pm$	0.0003	\\	
10725.41	&	-48507.3	$\pm$	1.3	&	3	&	-4.954	$\pm$	0.010	&	0.1793	$\pm$	0.0004	\\	
10730.40	&	-48507.8	$\pm$	0.7	&	9.8	&	-4.953	$\pm$	0.004	&	0.1792	$\pm$	0.0002	\\	
10731.42	&	-48509.7	$\pm$	1.1	&	4	&	-4.951	$\pm$	0.007	&	0.1781	$\pm$	0.0003	\\	
10732.39	&	-48509.6	$\pm$	1.4	&	9	&	-4.946	$\pm$	0.011	&	0.1793	$\pm$	0.0004	\\	
10749.36	&	-48505.9	$\pm$	0.8	&	9.2	&	-4.940	$\pm$	0.004	&	0.1789	$\pm$	0.0002	\\	
10753.34	&	-48509.4	$\pm$	1.1	&	12	&	-4.945	$\pm$	0.008	&	0.1777	$\pm$	0.0003	\\	
	
\hline																	
\end{tabular}															
\vspace{0.5em}	
\par
\captionsetup{width=0.6\textwidth}
\caption*{\textbf{Notes.} The $\rm \log R'_{HK}$ index was computed using YABI workflow implementation, while the H$\alpha$ activity index was extracted using the ACTIN 2 code.}								
\end{table}					

\clearpage
\section{Analysis:additional tables}

\renewcommand{\arraystretch}{1.3} 
\begin{table*}[ht!]		
\small					
\centering
\caption{TOI-4602 parameters from the transit and RV joint fit, obtained with one planet on a circular orbit model (M1 model).}					
\label{tab:join_fit_results}					
\begin{tabular}{lccc}					
\hline					
\hline					
\rule{0pt}{2.8ex}Parameter & Prior$^{(a)}$ & Value (circular model) & Value (eccentric model) \\					
\hline					
\multicolumn{3}{l}{\normalsize \rule{0pt}{2.8ex}Model parameters} \\[0.6ex]						
Orbital period P$_{orb}$ (days)	&	$\mathcal{N}$[3.98,0.2]	&	3.9812999$^{+0.0000047}_{-0.0000048}$ & 3.9812997$^{+0.0000048}_{-0.0000046}$	\\
Central time of the first transit T$_{0}$ (BJD - 2\,450\,000)	&	$\mathcal{U}$[10590, 10595]	&	10591.50097$^{+0.00066}_{-0.00073}$ & 10591.50080$^{+0.00070}_{-0.00076}$	\\
Scaled planetary radius R$_p$/R$_{\star}$	&	$\mathcal{U}$[0, 0.5]	&	$0.02083 \pm 0.00026$ & 	$0.02089 \pm 0.00076$\\
Impact parameter, $b$	&	$\mathcal{U}$[0, 2]	&	0.17$^{+0.16}_{-0.12}$ & 0.22$^{+0.20}_{-0.15}$	\\
Radial velocity semi-amplitude variation $K$ (m\,s$^{-1}$)	&	$\mathcal{U}$[0.01,30]	&	2.39$ \pm 0.40$ & 2.76$\pm$ 0.38	\\
\hline					
\multicolumn{3}{l}{\normalsize \rule{0pt}{2.8ex}Derived parameters} \\[0.6ex]					
Planetary radius (R$_J$)	&		&	0.219$^{+0.021}_{-0.020}$ & 0.220$\pm 0.020$\\
Planetary radius (R$_{\oplus}$)	&		&	2.45 $\pm 0.23$	& 2.46$\pm 0.23$\\
Planetary mass (M$_J$)	&		&	0.0174$^{+0.0029}_{-0.0030}$ & 0.0195$\pm$0.027	\\
Planetary mass (M$_{\oplus}$)	&		&
5.53$^{+0.93}_{-0.94}$	& 6.19$^{+0.86}_{-0.85}$\\
Eccentricity $e$	&  &	0 (fixed)	& 0.228$\pm$0.057\\
Scaled semi-major axis $a/R_{\star}$	&		&	8.90$^{+0.13}_{-0.35}$ & 9.22$^{+0.63}_{-0.65}$	\\
Semi-major axis $a$ (AU)	&		&	0.04747$^{+0.00035}_{-0.00036}$ &  0.04746$\pm$0.00035	\\
Argument of periastron $\omega_P$ (deg)	&	&	90 (fixed) &-157$^{+16}_{-21}$	\\
Orbital inclination $i$ (deg)	&		&	88.93$^{+0.74}_{-1.1}$ & 88.67$^{+0.89}_{-1.2}$	\\
Transit duration $T_{41}$ (days)	&		&	0.14364$\pm 0.00098$	& 0.135$^{+0.011}_{-0.010}$\\
Transit duration $T_{32}$ (days)	&		&	0.13744$^{+0.00095}_{-0.00096}$	& 0.129$\pm 0.010$\\
\hline					
\multicolumn{3}{l}{\normalsize \rule{0pt}{2.8ex}Calculated parameters} \\[0.6ex]						
Equilibrium temperature $T_{eq}$ (K)	&		&	1344$\pm$80	& 1321$\pm$78\\
Planetary density $\rho_P$ (g\,cm$^{-3}$)	&		&	2.18$\pm$0.77	& 2.41$\pm$0.82\\
Scale height $H$ (km)	&		&	538$\pm$140	& 476$\pm$114\\
$\delta_\lambda$ (ppm)	&		&	60$\pm$20 & 53$\pm$17	\\
Transmission Spectroscopy Metric (TSM)	&		&	140$\pm$54	& 125$\pm$46\\
Insolation (S$_{\oplus}$)	&		&	664$\pm$42 & 620$\pm$38	\\
\hline					
\multicolumn{3}{l}{\normalsize \rule{0pt}{2.8ex}Other system parameters} \\[0.6ex]						
Jitter term $\sigma_{HARPS-N}$ (m\,s$^{-1}$)	&	$\mathcal{U}$[0, 20]	&	1.58$^{+0.26}_{-0.23}$	& 1.33$^{+0.25}_{-0.21}$\\
Offset $\gamma_{HARPS-N}$  (m\,s$^{-1}$)	&	$\mathcal{U}$[-58514.0504, -38502.3977]	&	-48508.55$\pm 0.26$	& -48508.55$^{+0.25}_{-0.27}$\\
Jitter term $\sigma_{TESS_{Sect 43}}$ (ppm)	&	$\mathcal{U}$[0, 0.0329]	&	0.0002613$\pm 0.0000036$ & 	0.0002613$^{+0.0000037}_{-0.0000036}$\\
Jitter term $\sigma_{TESS_{Sect 44}}$ (ppm)	&	$\mathcal{U}$[0, 0.0328]	&	0.0002923$^{+0.0000036}_{-0.0000035}$ & 0.0002924$^{+0.0000036}_{-0.0000037}$	\\
Jitter term $\sigma_{TESS_{Sect 70}}$ (ppm)	&	$\mathcal{U}$[0, 0.0326]	&	0.0001863$^{+0.0000042}_{-0.0000041}$ & 0.0001864$^{+0.0000041}_{-0.0000042}$	\\
Jitter term $\sigma_{TESS_{Sect 71}}$ (ppm)	&	$\mathcal{U}$[0, 0.0341]	&	0.0002044$^{+0.0000041}_{-0.0000042}$ & 0.0002043 $\pm$ 0.0000042	\\
Jitter term $\sigma_{TESS_{Sect 86}}$ (ppm)	&	$\mathcal{U}$[0, 0.0374]	&	0.0002297$^{+0.0000044}_{-0.0000045}$ & 0.0002297$^{+0.0000045}_{-0.0000046}$	\\
Stellar density $\rho_\star$ ($\rho_\odot$)	&	$\mathcal{N}$[0.71, 0.16]	&	0.598$^{+0.026}_{-0.067}$ & 0.66$^{+0.15}_{-0.13}$	\\
Limb darkening $c_1$	&	$\mathcal{N}$[0.3876, 0.2]	&	0.39$ \pm 0.11$ & 0.39$\pm 0.11$	\\
Limb darkening $c_2$	&	$\mathcal{N}$[0.1413, 0.2]	&	0.08$^\pm 0.16$	& 0.08$\pm 0.15$\\
\hline					
\end{tabular}					
\vspace{0.5em}
\par
\captionsetup{width=0.98\textwidth}
\caption*{\textbf{Note.}  Parameter estimates and corresponding uncertainties are defined as the median and the 16th and 84th percentiles of the posterior distributions. $^{(a)} \mathcal{U}$[a, b] refers to uniform priors between a and b, $\mathcal{N}$[a, b] to Gaussian priors with median a and standard deviation b. $^{(b)}$ Parameter estimates and corresponding uncertainties are defined as the median and the 16th and 84th percentiles of the posterior distributions.}
\end{table*}					
\renewcommand{\arraystretch}{1} 

\clearpage
\twocolumn
\section{Transit time variations}

\begin{figure}[ht!]	
    \centering
    \includegraphics[width=1\columnwidth]{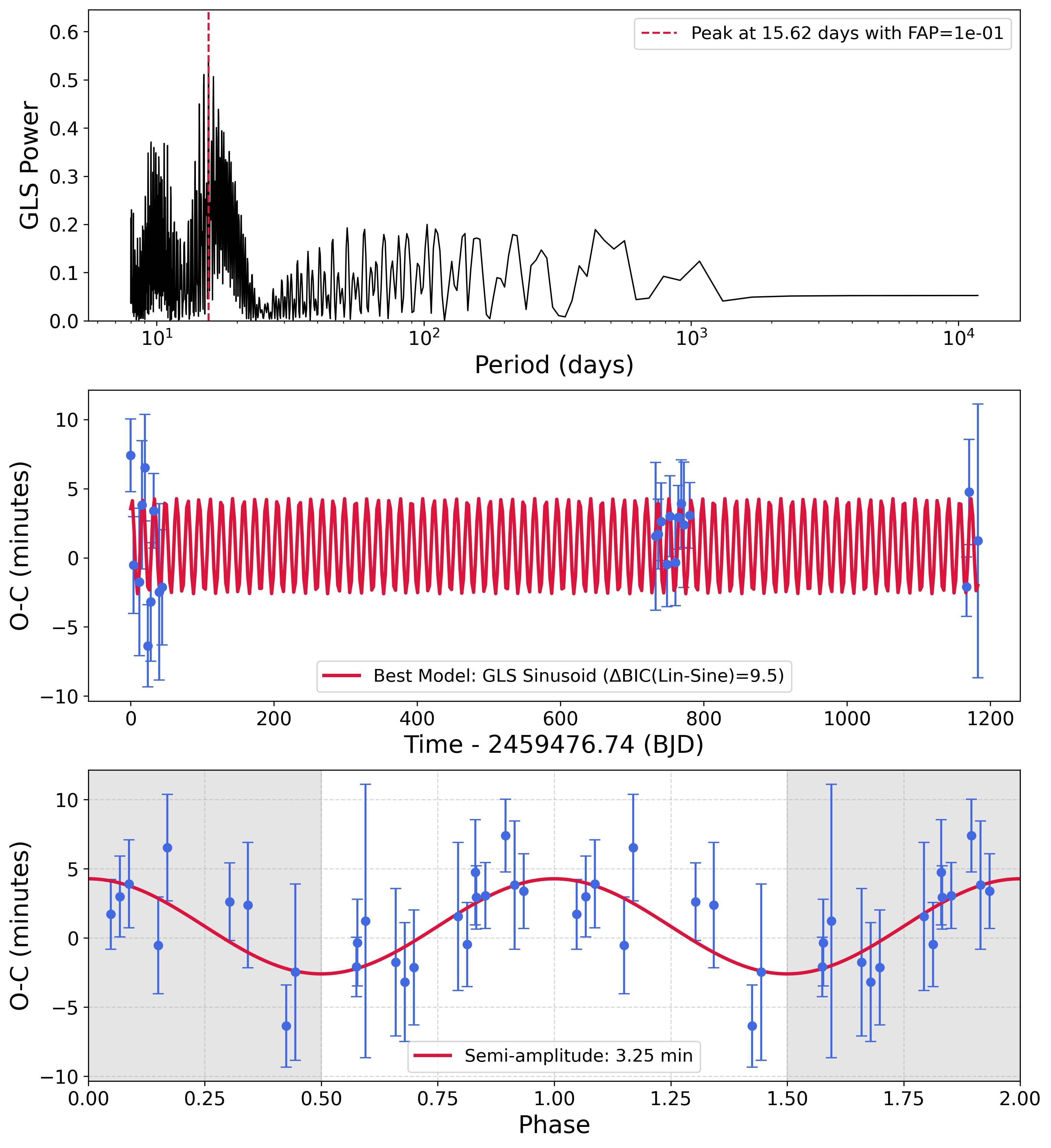}
    \caption{\textit{Top panel}: GLS of the Observed-Calculated (O-C) times of transits. \textit{Middle panel}: O-C times along with the best sinusoidal fitted model. \textit{Bottom panel}: O-C times in phase with the best fitted period.}
    \label{fig:TTV}
\end{figure}

\section{Detection sensitivity}\label{sec:completeness}

To assess our sensitivity to planetary companions across the parameter space, we performed an injection-recovery test combining both HARPS-N RVs and the PMa. In particular, we computed the detection probability on a logarithmic grid of planetary mass ($M_p \sin i$ for RV-only, $M_p$ for combined analysis), spanning from $10^{-2}\,M_{\rm{Jup}}$ to $20\,M_{\rm{Jup}}$, versus semi-major axis ($a$), from $10^{-2}$ au to $10$ au. For each cell of the grid, we simulated 200 synthetic planetary signals. The orbital parameters were drawn as follows: the period was determined from $a$ and the stellar mass; the time of conjunction and the argument of periastron were drawn from a uniform distribution; the orbital inclination ($i$) was drawn from a sinusoidal distribution (uniform in $\cos i$); and the eccentricity was drawn from a Beta distribution, following the statistical properties of known exoplanets \citep{Kipping2013}.

For the RV analysis, we adopted a model comparison approach based on the BIC. For each simulated planet, we generated a synthetic RV time series at the epochs of the observed data, adding Gaussian noise scaled by the instrumental jitter ($\sigma_{\rm HARPS-N}$) from Table\,\ref{tab:join_fit_results}. We then fitted the synthetic data with three different models:
\begin{enumerate}
    \item A constant model (no planet, $k=1$ parameter);
    \item A polynomial trend model (linear or quadratic, $k=2,3$) to account for long-period companions;
    \item A full Keplerian orbit model ($k=6$).
\end{enumerate}

A planet was considered "detected" via RVs if the Keplerian model provided significantly better evidence than the polynomial or constant baseline models. Specifically, we required a difference $\Delta \text{BIC} > 10$ \citep{Kass01061995}, where $\text{BIC} = \chi^2 + k \ln(N_{\rm obs})$. Furthermore, planets inducing strong linear or quadratic trends (where the orbital period is significantly longer than the observational baseline) were also flagged as detected if the polynomial model was favored over the constant model.

To evaluate the sensitivity to wide-orbit companions, we further utilized the PMa measured between the \textit{Hipparcos} and \textit{Gaia} eDR3 catalogues \citep{Brandt2018}. The PMa vector corresponds to the difference between the long-term proper motion vector (approximating the centre-of-mass motion) and the short-term \textit{Gaia} proper motion. We defined the astrometric detection threshold at a $3\sigma$ level. For each $a$, we calculated the minimum planetary mass $M_{\rm min}$ required to induce a tangential velocity anomaly $\Delta v_T$ equal to three times the measurement error ($\sigma_{\Delta v_T}$) reported in the HGCA. This calculation accounts for the orbital smearing effect due to the finite observing baselines of the two missions, following the formulation by \cite{Kervella2019}. A simulated planet was considered astrometrically detected if its mass $M_p > M_{\rm min}(a)$.

\begin{figure}[hb!]	
    \centering
    \includegraphics[width=1\columnwidth]{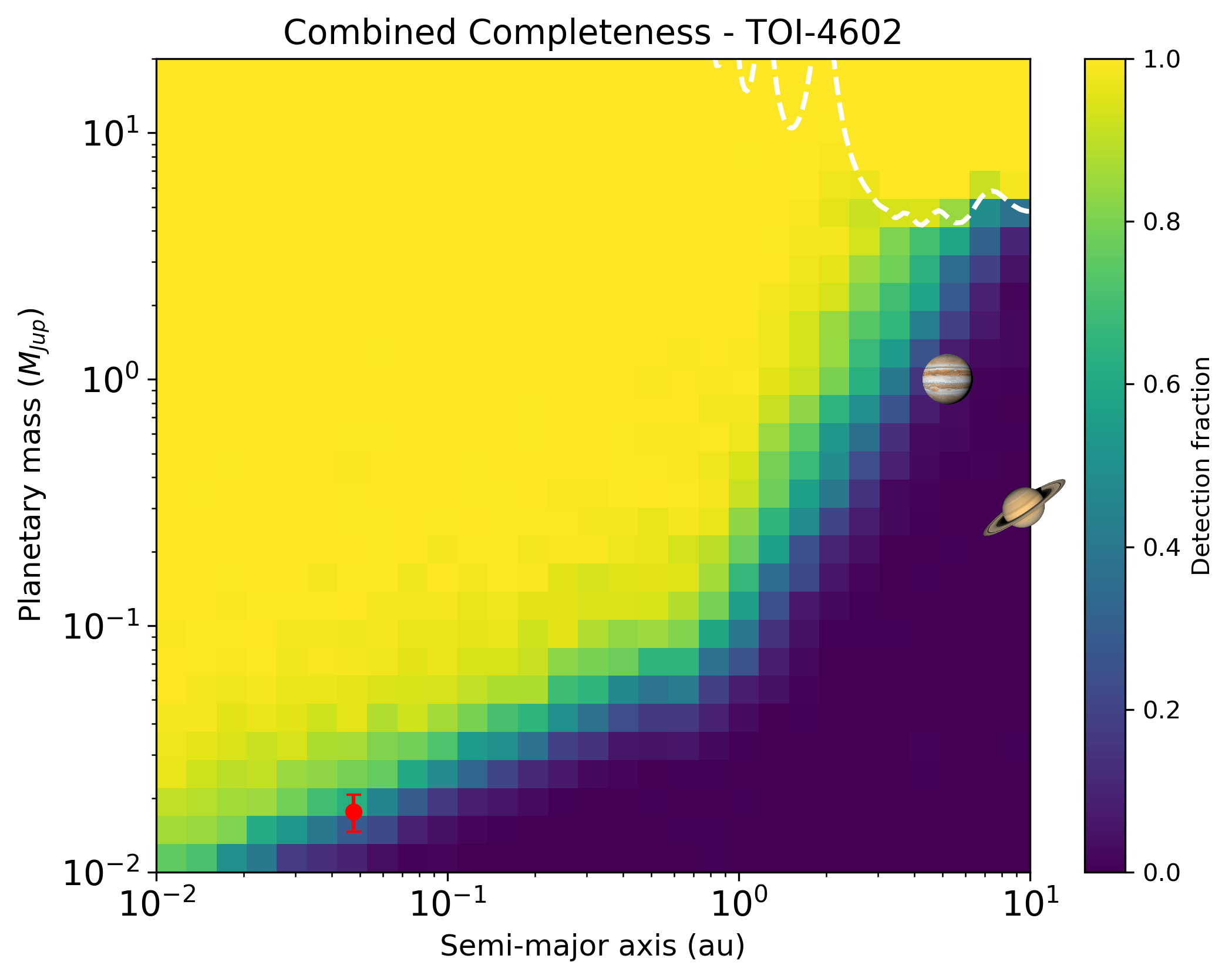}
    \caption{Detection completeness map for TOI-4602 in the mass versus semi-major axis plane, combining RV and PMa constraints. The colour scale represents the detection fraction of simulated planetary companions. 
    The white dashed line indicates the astrometric detection limit, corresponding to a PMa signal of $3\sigma$ based on the \textit{Hipparcos}-\textit{Gaia} eDR3 baseline. Regions above this line are accessible via astrometry, while the inner regions are probed by RVs. The red circle marks the positions of TOI-4602\,b. For context, Jupiter and Saturn are shown as icons.}
    \label{fig:completeness}
\end{figure}

The final completeness map (Fig.\,\ref{fig:completeness}) represents the union of the two detection methods. A simulated planet in a given grid cell was considered recovered if it satisfied \textit{either} the RV detection criterion ($\Delta \text{BIC} > 10$) or the PMa detection criterion (signal $> 3\sigma$). The detection fraction for each cell is defined as $N_{\rm det} / N_{\rm sim}$. Currently, the data is sensitive to Jupiter-mass planets only up to $\sim1$ au.

\clearpage
\section{Mass-radius diagram and internal structure: additional figures and tables}
\begin{figure}[H]
	\centering
	\includegraphics[width=1\columnwidth]{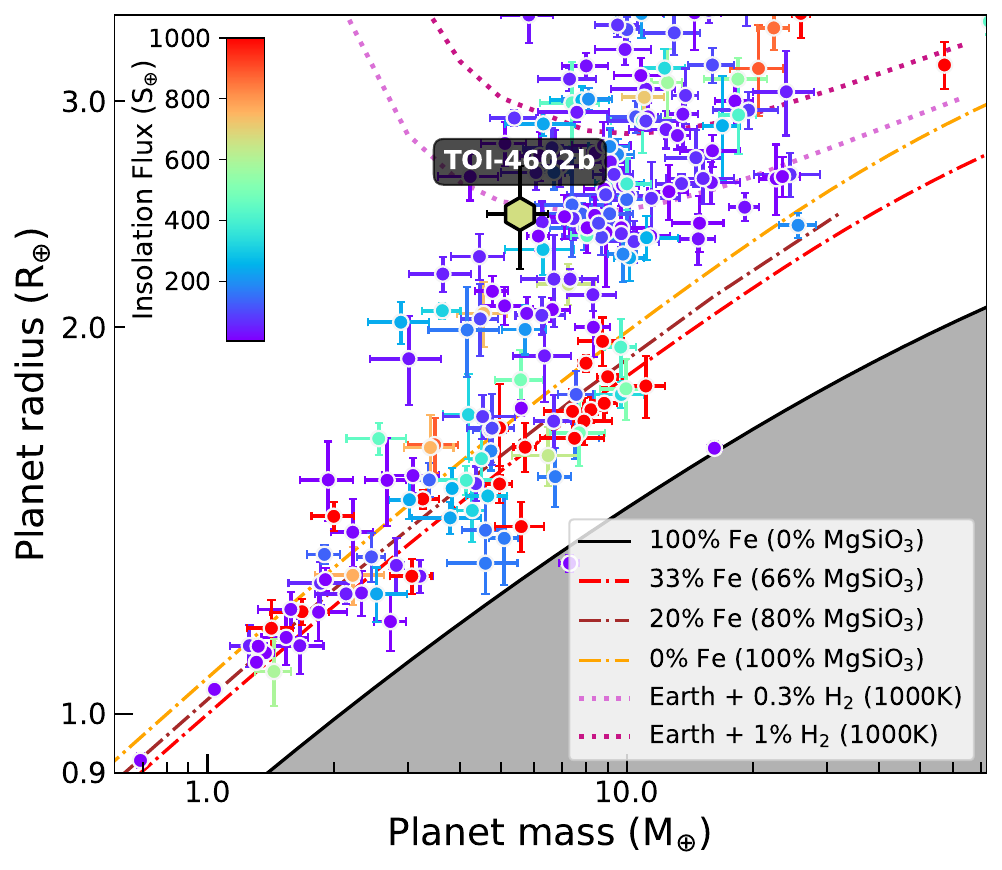}
	\caption{Mass-radius diagram of exoplanets with mass and radius derived with precision better than 20\% and 10\%, respectively.  The colour indicates the stellar insolation in Earth units (see colour bar in the upper left corner). The mass and radius of TOI-4602\,b, as derived in this work, are highlighted on the diagram.  Theoretical mass-radius curves, computed assuming a surface temperature of 1000 K, are overplotted for reference. The curve corresponding to a 20\% Fe core mass fraction is taken from \citet{Zeng2016ApJ...819..127Z}, while the remaining composition models are from \citet{Zeng2019PNAS..116.9723Z}. From top to bottom, the curves correspond to: Earth-like rocky cores (32.5\% Fe + 67.5\% MgSiO$_3$) with a 0.3\% and 1\% H$_2$ envelopes, and rocky compositions with varying silicates fractions (100\% Fe, 33\% Fe + 66\% MgSiO$_3$, 20\% Fe + 80\% MgSiO$_3$, 100\% MgSiO$_3$). This plot has been generated with \texttt{mr-plotter} \citep[][https://github.com/castro-gzlz/mr-plotter/]{Castro-Gonzalez2023A&A...675A..52C}.}
		\label{fig:massa-radius-diagram}
\end{figure}

\begin{table}[hb!]\footnotesize
\centering
\caption{Prior and posterior parameter distributions for the interior characterization assuming an Earth-like rocky composition.}
\label{tab:priors_posteriors_combined}
\renewcommand{\arraystretch}{1.2}
\setlength{\tabcolsep}{3pt} 
\begin{tabular}{lccc}
\hline\hline
Parameter 
& Prior 
& \makecell{Posterior \\ (circular model)} 
& \makecell{Posterior \\ (eccentric model)} \\
\hline \\[-10pt] 
$M_\mathrm{atm}$ 
& $\mathcal{U}(5e^{-3} - 1e^{-1})\,M_p$ 
& $0.05 \pm 0.03\,M_{\oplus}$ 
& $0.05 \pm 0.03\,M_{\oplus}$ \\
$M_\mathrm{core+mantle}$ 
& $\mathcal{N}(M_p,\sigma^2_{M_p})$ 
& $5.62^{+0.86}_{-0.89}\,M_{\oplus}$ 
& $6.24^{+0.81}_{-0.87}\,M_{\oplus}$ \\
$Z_\mathrm{env}$ 
& $\mathcal{U}(0.0 - 1.0)$ 
& $0.61^{+0.24}_{-0.26}$ 
& $0.60^{+0.24}_{-0.27}$ \\
\bottomrule
\end{tabular}
\par
\captionsetup{width=0.49\textwidth}
\caption*{\textbf{Notes.} Priors are identical for the circular and eccentric models except for the value of $M_p$. Posterior values correspond to the circular and eccentric solutions shown in Fig.~\ref{fig:inversion}.}			
\end{table}

\begin{figure}[H]
    \centering
    \includegraphics[width=0.8\columnwidth]       {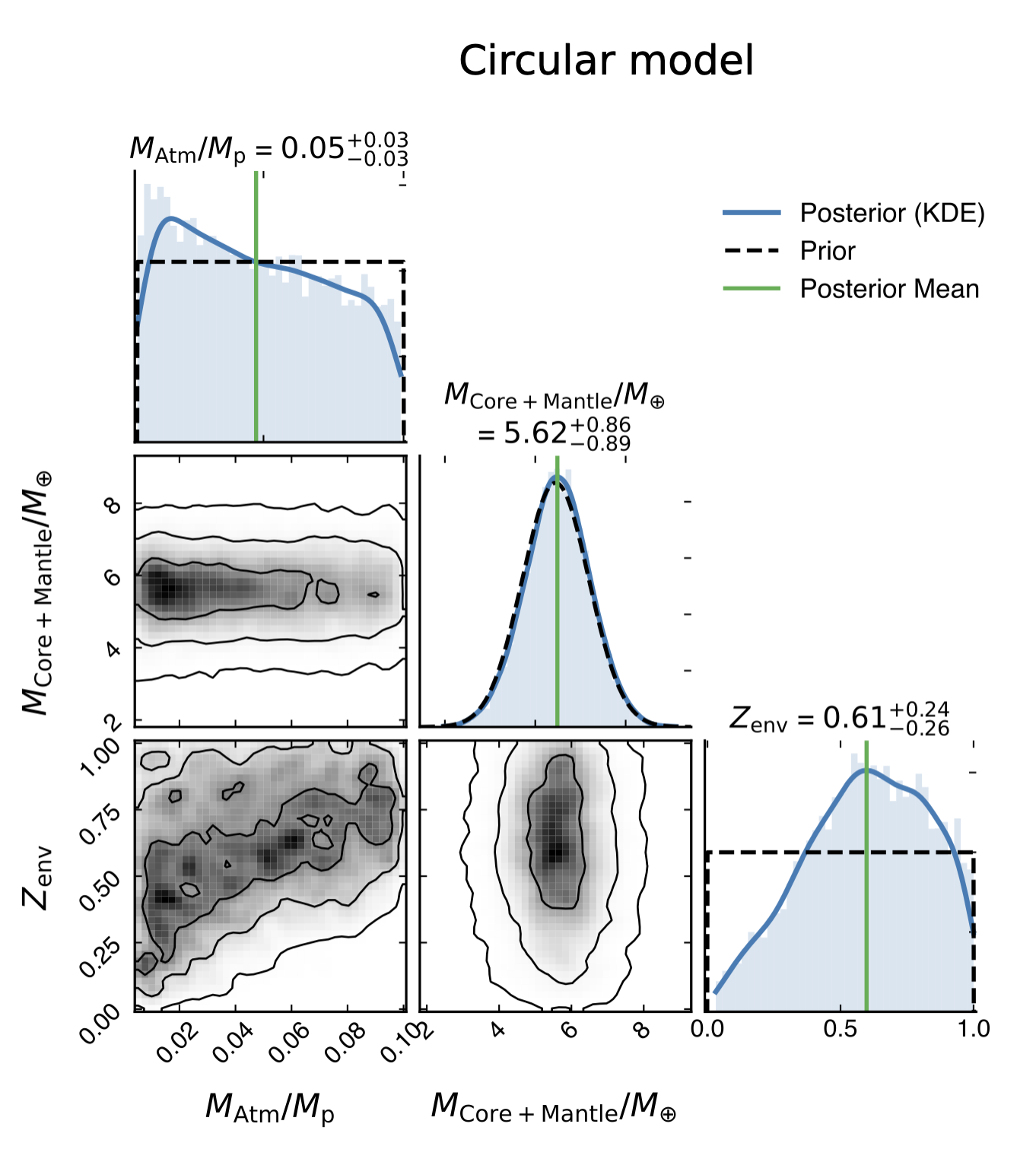}\\
    \includegraphics[width=0.8\columnwidth]{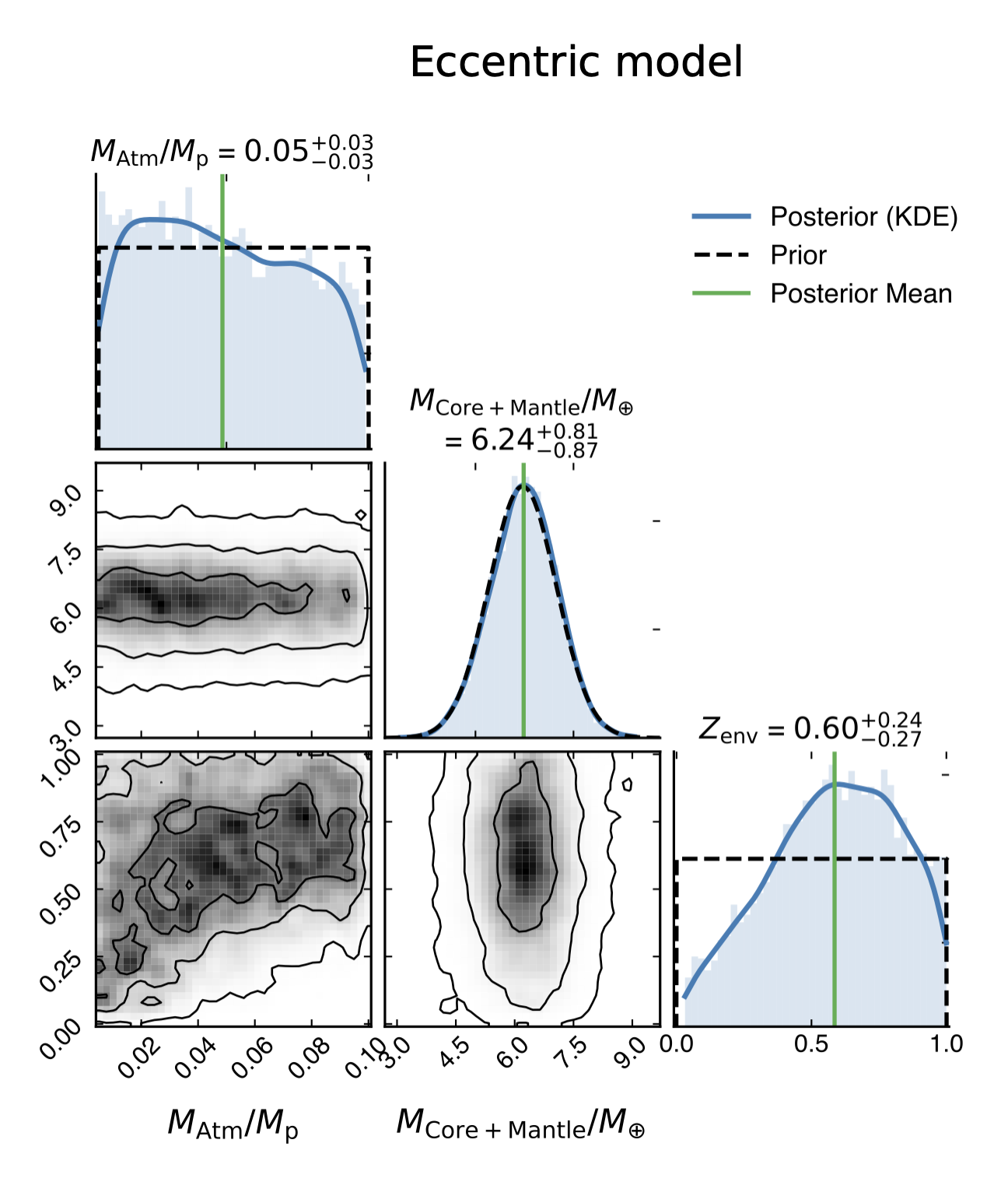}

    \caption{Marginalized posterior distribution of interior parameters in 1-D and 2-D. The shown posterior is calculated assuming an enriched H/He envelope above an Earth-like interior and is constrained by planet mass, radius, equilibrium temperature, and age. }
    \label{fig:inversion}
\end{figure}

\onecolumn
\clearpage

\section{Photevaporation and long-term evolution: additional figures} 
\vspace{-0.5cm}
\begin{figure}[H]
\centering
\begin{tabular}{@{}c@{}c@{}c@{}}
   \hspace{-0.5cm} \includegraphics[width=6.7cm]{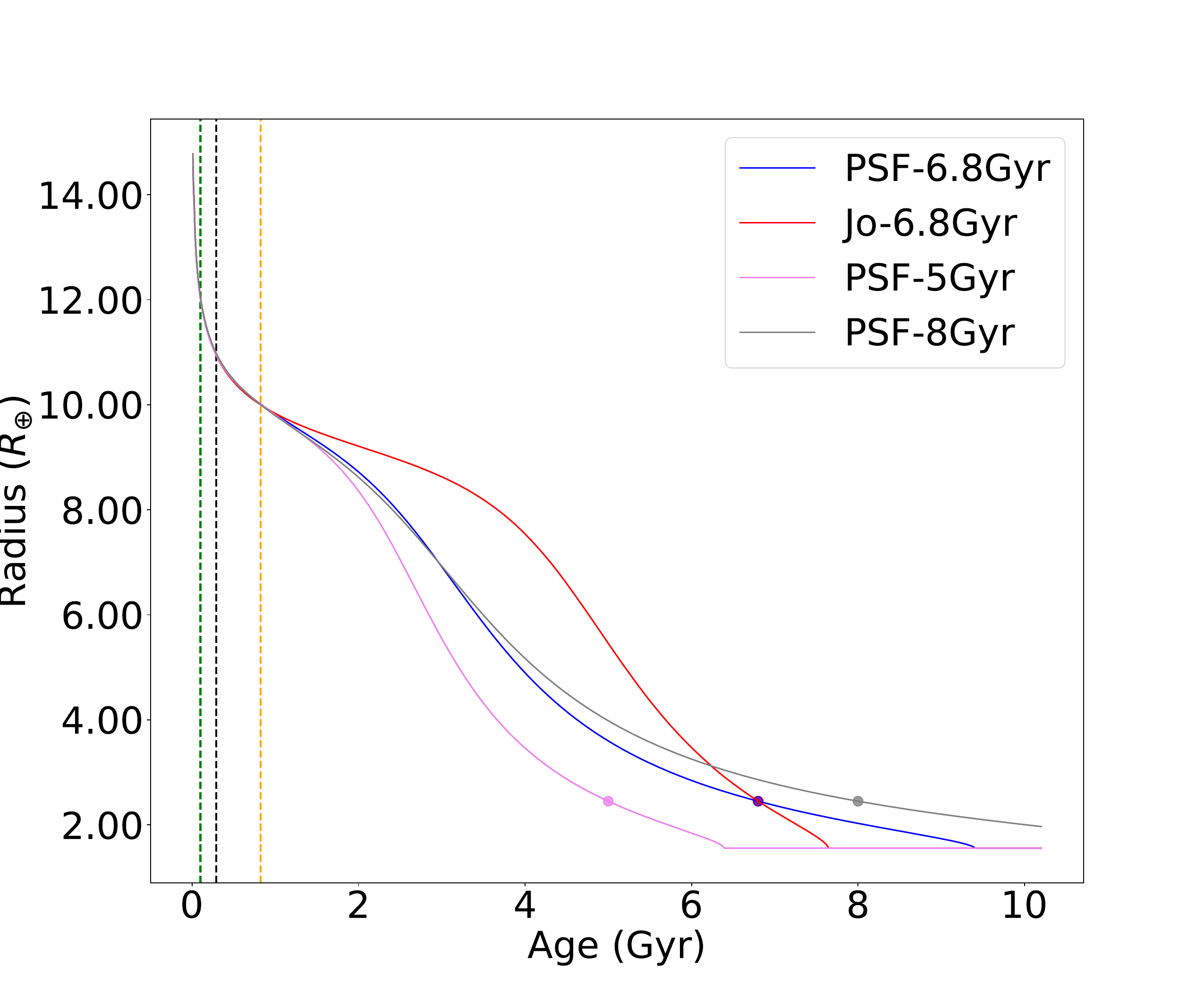} & \hspace{-0.5cm}
    \includegraphics[width=6.7cm]{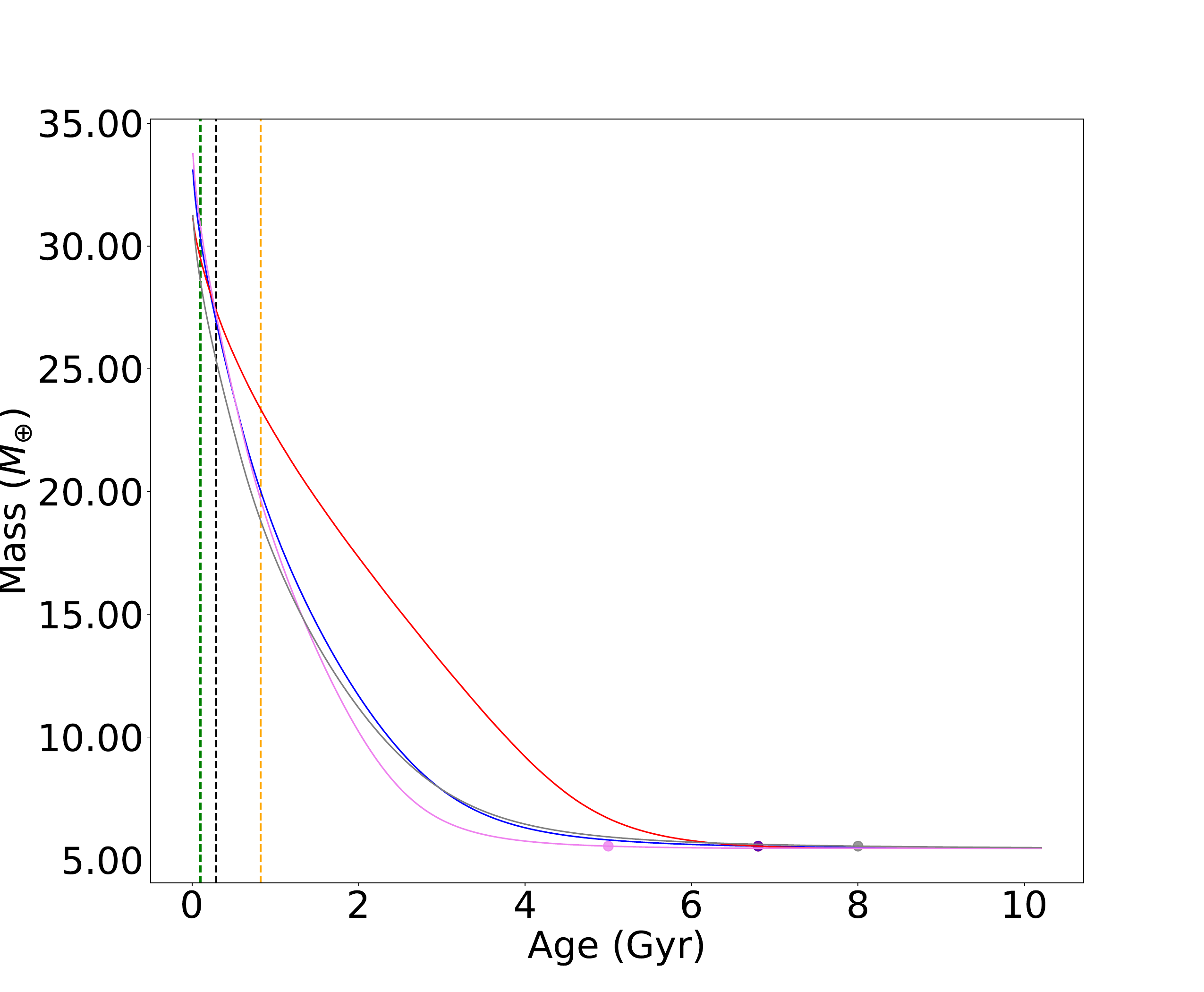} &  \hspace{-0.5cm}
    \includegraphics[width=6.7cm]{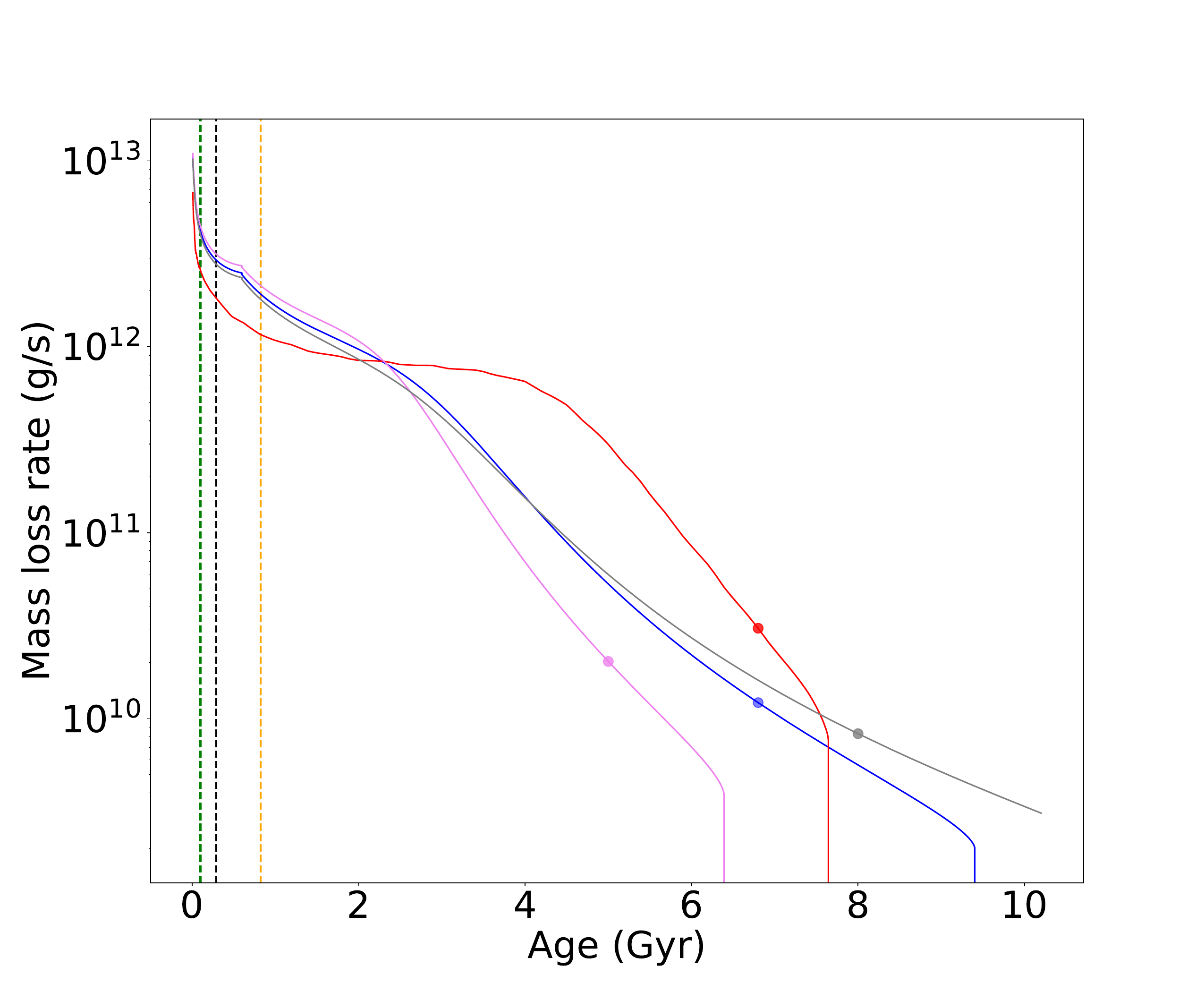} \\
\end{tabular}
\caption{Temporal evolution of planetary parameters of TOI-4602\,b. The left, middle and right panels show the evolution of  radius, mass, and mass-loss rate, respectively. The colours refers to  different high-energy evolutionary tracks as described in the text. The coloured circles indicate the planet's current position, which depends on the age assumed in the model. In the left and middle panels, the red and blue circles coincide. Vertical dashed lines mark the limit of validity for analytical models, beyond which data are extrapolated to 10 Myr. These include the \citet{LopFor14} envelope radius equation (orange), the mass loss rate ATES expression (black), and the \citet{Penz+2008} X-ray luminosity equation (green) ( \citealt{Johnstone+2021} track is valid beyond these thresholds).}
\label{fig:evap}
\end{figure}
\vspace{-0.8cm}

\begin{figure}[H]
\centering
\begin{tabular}{@{}c@{}c@{}c@{}}
   \hspace{-0.5cm} \includegraphics[width=6.7cm]{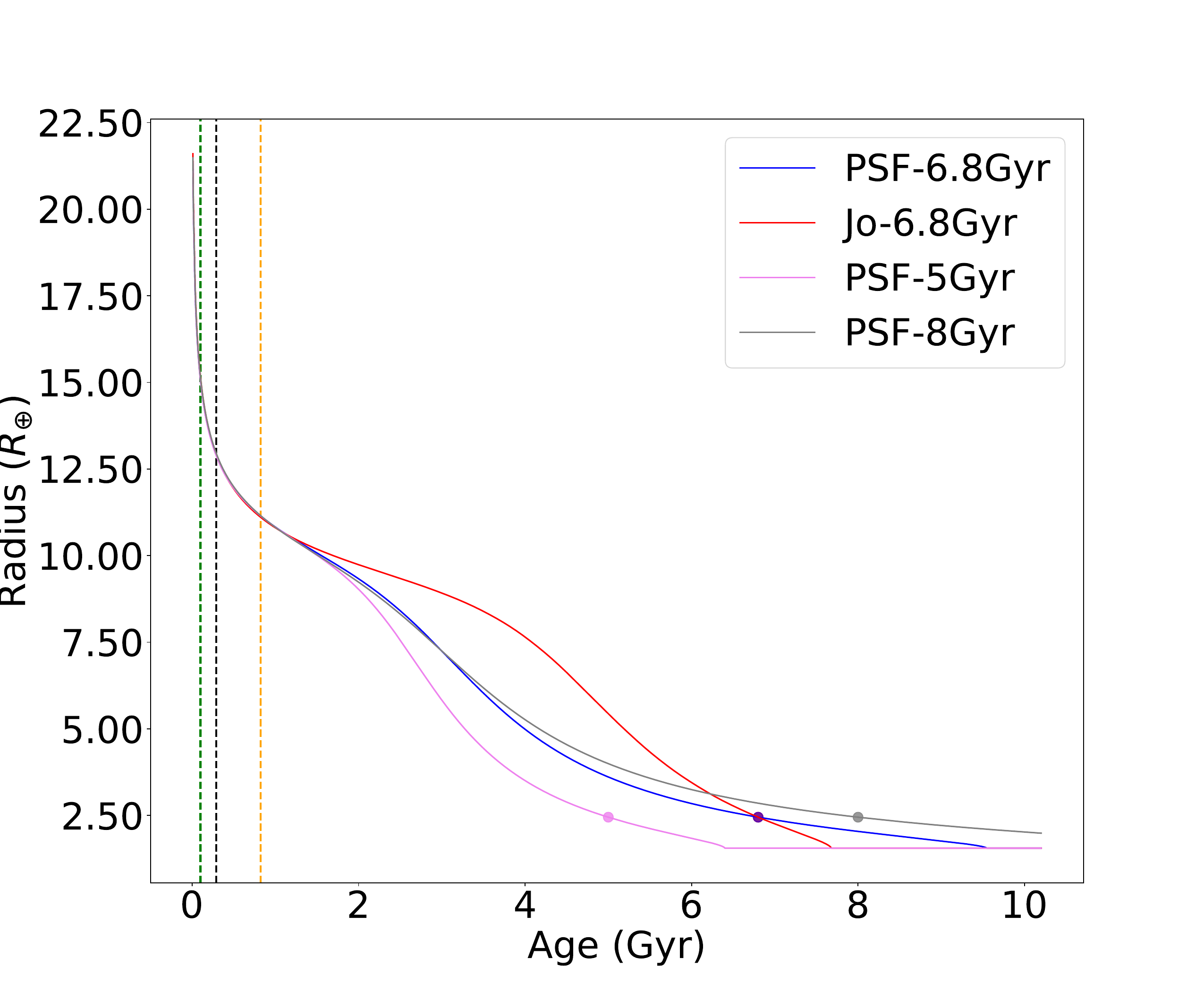} & \hspace{-0.5cm}
    \includegraphics[width=6.7cm]{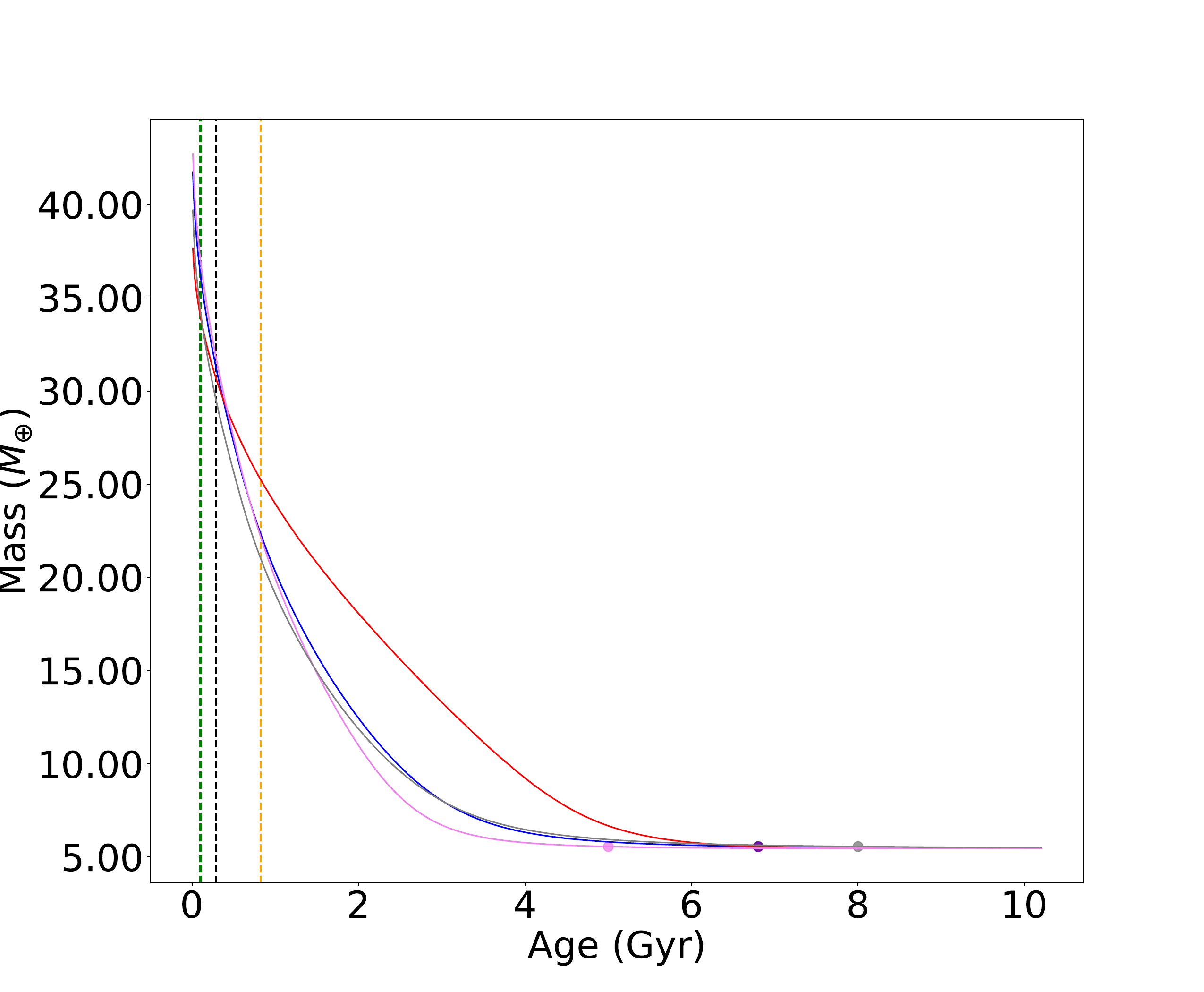} & \hspace{-0.5cm}
    \includegraphics[width=6.7cm]{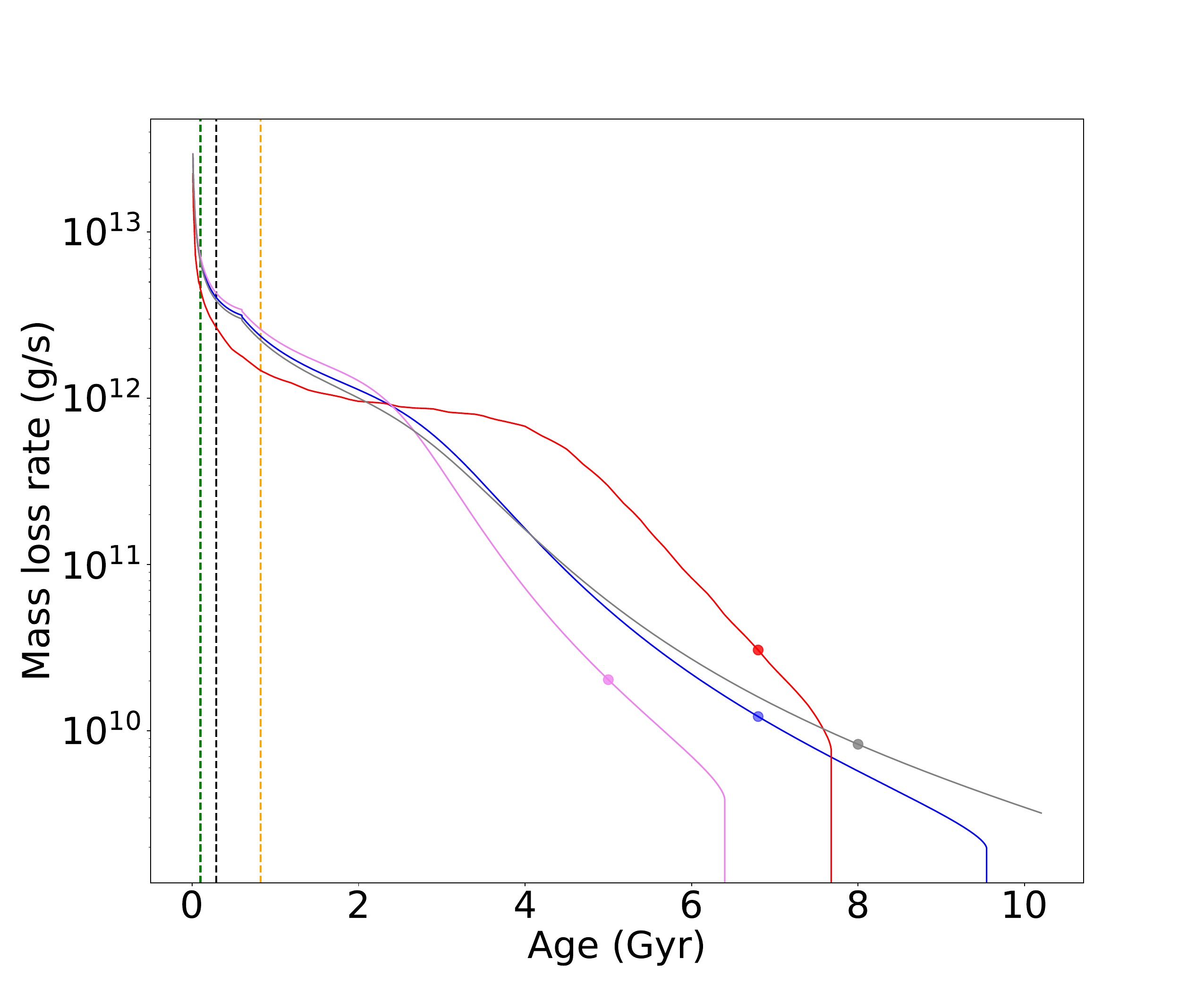} \\
\end{tabular}
\caption{The same of Fig. \ref{fig:evap} for the enhanced metallicity composition of the planet atmosphere (see Sect. \ref{sec:fotoevaporazione}).}
\label{fig:evap50x}
\end{figure}

\section{Atmospheric simulations: additional figures}
\begin{figure}[H]
    \centering
    \includegraphics[width=0.49\columnwidth]{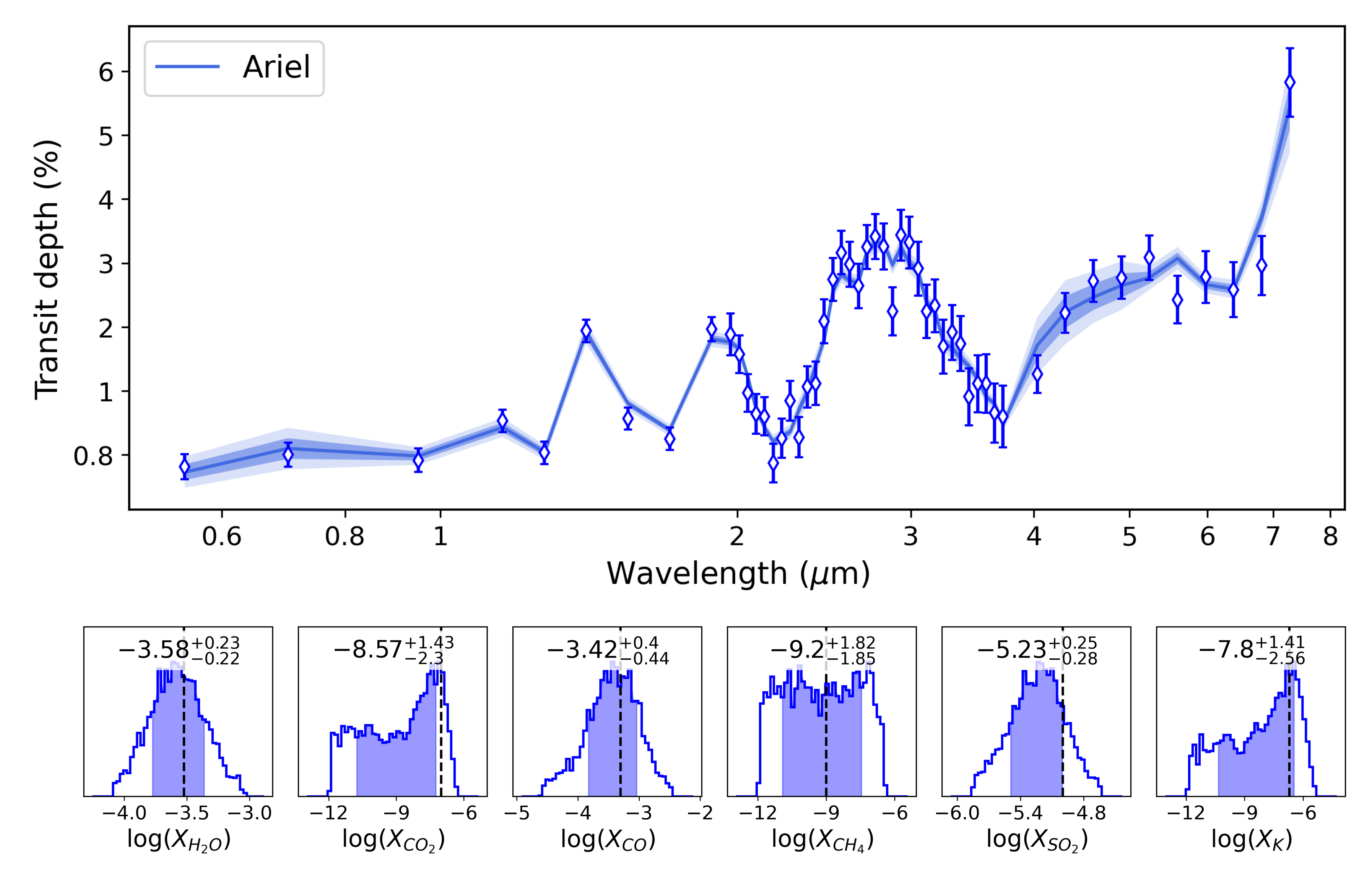}
    \includegraphics[width=0.49\columnwidth]{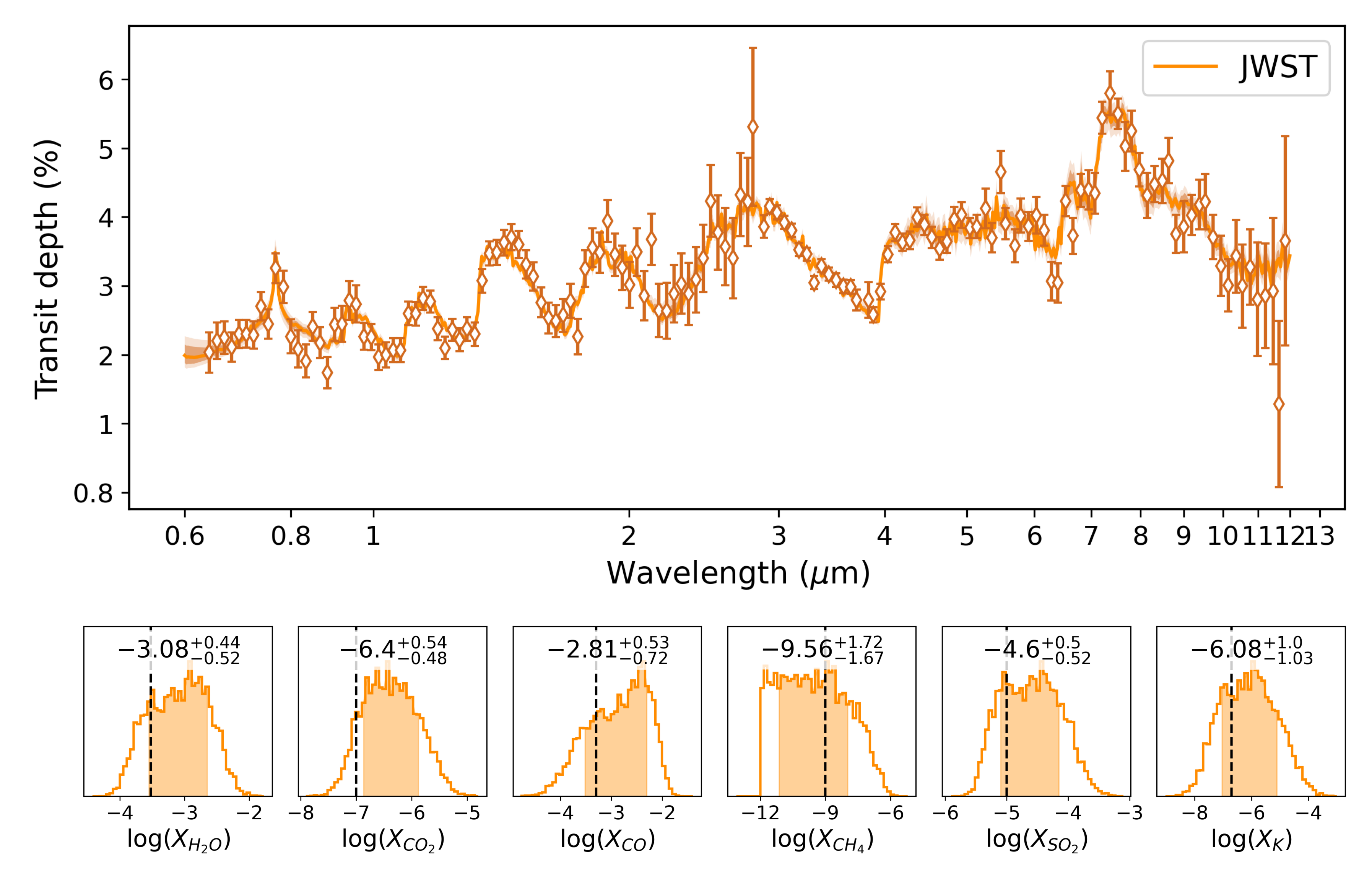}

    \caption{Simulated spectra of TOI-4602\,b as observed by Ariel (left panels)
    and JWST (right panels). Bottom panels show the posterior distribution of the VMRs of the molecules for each simulation, respectively. The shaded areas represent the 1$\sigma$ uncertainty, while the values report the media and 1$\sigma$ span of the posteriors. Black dashed lines highlight the VMRs of the input model. }
    \label{fig:atmo_simulations}
\end{figure}
\end{document}